\documentclass[letterpaper,twocolumn,10pt]{article}
\usepackage{usenix}

\usepackage{epsfig,amsmath,amsfonts,epsfig,multirow,makecell,caption,soul,csquotes,color,wrapfig,subcaption,mathtools,bm,spverbatim,booktabs,tcolorbox,diagbox}
\usepackage[e]{esvect}

\usepackage{caption}
\newtcolorbox[auto counter]{mtbox}[1]{left=0.25mm, right=0.25mm, top=0.25mm, bottom=0.25mm, sharp corners, colframe=blue!50!black, boxrule=0.5pt, title={#1~\thetcbcounter}, fonttitle=\bfseries, coltitle=blue!50!black, attach title to upper={\ --\ }}

\makeatletter
\newif\if@restonecol
\makeatother

\usepackage[boxed, ruled, vlined, algo2e, linesnumbered]{algorithm2e}
\SetKwRepeat{Do}{do}{while}

\newenvironment{changemargin}[2]{\begin{list}{}{
	\setlength{\topsep}{0pt}\setlength{\leftmargin}{0pt}
	\setlength{\rightmargin}{0pt}
	\setlength{\listparindent}{\parindent}
	\setlength{\itemindent}{\parindent}
	\setlength{\parsep}{0pt plus 1pt}
	\addtolength{\leftmargin}{#1}\addtolength{\rightmargin}{#2}
	}\item}
	{\end{list}}

\newenvironment{mitemize}{
	\begin{changemargin}{-10pt}{-0cm}
	\vspace{-10pt}
	\hspace{0pt}
	\begin{itemize}
	\setlength{\itemsep}{3pt}}
	{\end{itemize}
	\vspace{2pt}
	\end{changemargin}}

\newcommand{\sboth}[3]{{#1}_{\scaleobj{0.8}{#2}}^{\scaleobj{0.8}{#3}}}

\usepackage[first=0,last=9]{lcg}
\usepackage{colortbl}
\definecolor{Gray}{gray}{0.8}

\usepackage{hyperref}

\newcommand{\msec}[1]{\S\,\ref{#1}}
\newcommand{\mref}[1]{\,\ref{#1}}
\newcommand{\meq}[1]{Eqn\,(\ref{#1})}
\newcommand{\mcite}[1]{\cite{#1}}

\newcommand{\meg}{{\em e.g.}\xspace}

\usepackage{scalerel}[2016/12/29]

\makeatletter
\providecommand{\leadsfrom}{%
  \mathrel{\mathpalette\reflect@squig\relax}%
}
\newcommand{\reflect@squig}[2]{%
  \reflectbox{$\m@th#1\leadsto$}%
}
\makeatother

\def\eqref#1{equation~\ref{#1}}

\def\1{\bm{1}}

\DeclareMathAlphabet{\mathsfit}{\encodingdefault}{\sfdefault}{m}{sl}
\SetMathAlphabet{\mathsfit}{bold}{\encodingdefault}{\sfdefault}{bx}{n}

\def\gD{{\mathcal{D}}}

\def\gL{{\mathcal{L}}}

\def\gN{{\mathcal{N}}}

\def\sE{{\mathbb{E}}}

\usepackage{microtype}
\usepackage{graphicx}
\usepackage{booktabs} 
\usepackage{amsfonts}
\usepackage{algorithm}
\usepackage{algorithm2e,amsmath,amssymb}
\usepackage{caption}

\makeatletter
\newcommand\footnoteref[1]{\protected@xdef\@thefnmark{\ref{#1}}\@footnotemark}
\makeatother

\newcommand{\rev}[1]{#1\xspace}

\usepackage{mathtools}
\usepackage{amsmath}
\usepackage{amsthm}
\usepackage{titlesec}

\theoremstyle{plain}
\newcommand{\attack}{{\sc Trl}\xspace}

\usepackage[textsize=tiny]{todonotes}

\begin{document}


\def\thetitle{\rev{On the Difficulty of Defending Contrastive Learning against Backdoor Attacks}}

\title{\thetitle }

\author{
{\rm Changjiang  Li}$^\star$ \quad {\rm Ren Pang}$^\dagger$ \quad {\rm Bochuan Cao}$^\dagger$ \quad {\rm Zhaohan Xi}$^\dagger$ \quad {\rm Jinghui Chen}$^\dagger$ \quad \\ {\rm Shouling Ji}$^\ddagger$ \quad {\rm Ting Wang}$^\star$\\
$^\star$ Stony Brook University \quad  
$^\dagger$Pennsylvania State University \quad  $^\ddagger$Zhejiang University \\
{\tt\small meet.cjli@gmail.com, \{rbp5354, bccao, zxx5113, jzc5917\}@psu.edu} \\
{\tt\small sji@zju.edu.cn,} 
{\tt\small inbox.ting@gmail.com}
} 

\maketitle

\begin{abstract}
Recent studies have shown that contrastive learning, like supervised learning, is highly vulnerable to backdoor attacks wherein malicious functions are injected into target models, only to be activated by specific triggers. However, thus far it remains under-explored how contrastive backdoor attacks fundamentally differ from their supervised counterparts, which impedes the development of effective defenses against the emerging threat. 

This work represents a solid step toward answering this critical question. Specifically, we define \attack\footnote{\label{note1}\rev{\attack: \underline{TR}ojan \underline{L}earning}}, a unified framework that encompasses both supervised and contrastive backdoor attacks. Through the lens of \attack, we uncover that the two types of attacks operate through distinctive mechanisms: in supervised attacks, the learning of benign and backdoor tasks tends to occur independently, while in contrastive attacks, the two tasks are deeply intertwined both in their representations and throughout their learning processes. This distinction leads to the disparate learning dynamics and feature distributions of supervised and contrastive attacks. More importantly, we reveal that the specificities of contrastive backdoor attacks entail important implications from a defense perspective: existing defenses for supervised attacks are often inadequate and not easily retrofitted to contrastive attacks. We also explore several alternative defenses and discuss their potential challenges. Our findings highlight the need for defenses tailored to the specificities of contrastive backdoor attacks, pointing to promising directions for future research.
\end{abstract}

\section{Introduction}

As an emerging machine learning paradigm, contrastive learning (CL) has gained significant advances recently\mcite{chen:2020:simple,grill:2020:bootstrap,newell2020useful,chen:2021:exploring,chen:2020:improved,he:2020:momentum}. Without requiring data labeling, CL is able to learn high-quality representations of complex data and enable a range of downstream tasks. Intuitively, CL performs representation learning by aligning the representations of the same input under varying data augmentations while separating the representations of different inputs. In various tasks, CL achieves performance comparable to supervised learning (SL)\mcite{grill:2020:bootstrap}. Many IT giants have unveiled their CL-based products and services\mcite{ramesh2022hierarchical, dalle2}. 

The surging popularity of CL also raises significant security concerns. Backdoor attacks represent one major threat, which injects malicious ``backdoors'' into target models, only to be activated by specific ``triggers''. For example, a backdoored model may misclassify trigger-embedded inputs to a target class while functioning normally on clean inputs\mcite{badnet}. Backdoor attacks are of special interest to CL. As CL-trained models are subsequently used in various downstream tasks, such attacks can cause widespread damage. Despite a plethora of backdoor attacks against SL\mcite{badnet, brendel2017decision, imc, trojannn, latent-backdoor}, due to the absence of labels, supervised backdoor attacks are often inapplicable to CL directly. Recent work has explored new ways of injecting backdoors into CL-trained models\mcite{saha:2021:backdoor,li2022demystifying,liu:2022:poisonedencoder}. For example, SSLBackdoor\mcite{saha:2021:backdoor} extends BadNets\mcite{badnet} by only poisoning inputs in the target class; PoisonedEncoder\mcite{liu:2022:poisonedencoder} generates poisoning data by randomly combining target inputs with reference inputs; CTRL\mcite{li2022demystifying} defines triggers as specific perturbations in the spectral domain of given inputs. While these studies show empirically that CL is also highly vulnerable to backdoor attacks, a set of key questions remain unexplored:
\begin{mitemize}
    \item Q1 -- How do contrastive backdoor attacks fundamentally differ from their supervised counterparts?
    \item Q2 -- What are the implications of their distinctions from a defense perspective?
    \item Q3 -- Is it feasible to retrofit defenses against supervised backdoor attacks to contrastive attacks?
\end{mitemize}

Answering these questions is critical for both (i) understanding the vulnerabilities of contrastive learning and (ii) developing effective defenses against the emerging threat.

\vspace{2pt}
{\bf Our work.}
This work represents a solid step toward answering these critical questions.

\vspace{2pt}
A1 -- We first define \attack\footnoteref{note1}, a unified framework that encompasses both supervised and contrastive attacks. Through the lens of \attack, we uncover that the two types of attacks operate through distinct mechanisms. In supervised attacks, the learning of benign and backdoor tasks occurs independently at the data level, focusing on learning trigger features {\it only} from the poisoning data and learning semantic features {\it only} from the clean data, while in contrastive attacks, the learning of benign and backdoor tasks is deeply intertwined not only in their representations but also throughout their learning processes.

\vspace{2pt}
A2 -- We then show that the disparate mechanisms of supervised and contrastive attacks lead to their distinctive learning dynamics and feature distributions. For instance, in supervised attacks, the loss of poisoning data often drops much faster than that of clean data, as learning trigger features often represents a simpler task than learning semantic features. In contrastive attacks, as the learning of benign and backdoor tasks is intertwined in the poisoning data, the loss of poisoning data often decreases at a rate similar to the clean data.

\vspace{2pt}
A3 -- More importantly, we reveal that the specificities of contrastive backdoor attacks entail unique challenges from a defense perspective. Defenses against supervised attacks that attempt to segregate poisoning data based on learning dynamics and feature distributions tend to fail. Also, defenses applied on the end-to-end model in the downstream task are often ineffective. Finally, we explore promising alternative defenses (e.g., data-free pruning and density-based filtering) and discuss their potential challenges.

\vspace{2pt}
{\bf Our contributions.}
To our best knowledge, this work is the first to systematically investigate the fundamental differences between supervised and contrastive backdoor attacks from a defense perspective. Our key contributions can be summarized as follows:
\begin{mitemize}
\item For the first time, we uncover the distinctive mechanisms underlying supervised and contrastive backdoor attacks.

\item We highlight the significant implications of such differences from a defense standpoint and reveal the inadequacy of existing defenses against contrastive backdoor attacks.

\item We explore promising alternative defense strategies and discuss their potential challenges, pointing to several directions for future research.
\end{mitemize}

We believe our findings shed new light on developing more robust contrastive learning techniques.

\section{Preliminaries}
\label{sec:background}

We first introduce the fundamental concepts used in the paper.

\subsection{Contrastive Learning}
A predictive model $h$ (parameterized by $\theta$) typically comprises two parts $h = g \circ f$ where an encoder $f$ extracts latent representations (i.e., features) from inputs while a classifier $g$ maps such features to output classes. Supervised learning optimizes parameters $\theta$ using a training set $\gD$ with labeled instances $(x, y)$ and a loss function $\sE_{(x,y) \in \gD} \,\ell(h_\theta(x), y)$, such as cross-entropy between $h_\theta(x)$ and $y$.

However, supervised learning is inapplicable when data labeling is scarce or expensive. Recently, contrastive learning (CL) has emerged as an alternative, which leverages the supervisory signals from the data itself to train $f$ that is then combined with $g$ and fine-tuned using weak supervision. Typically, CL performs representation learning by aligning the features of the same input under varying augmentations (i.e., ``positive pairs'') and separating the features of different inputs (i.e., ``negative pairs'').

A variety of CL methods (e.g., SimCLR\mcite{chen:2020:simple}, BYOL\mcite{grill:2020:bootstrap}, 
MoCo\mcite{he:2020:momentum}) have been proposed and garnered significant attention. For instance, SimCLR maximizes the similarity of positive pairs relative to the similarity of negative pairs. Specifically, for each input $x$, a pair of its augmented views $(x, x^+)$ forms a positive pair, while a set of augmented views of other inputs $\gN_x$ forms its negative inputs. The contrastive loss is defined by the InfoNCE loss\mcite{infonce}:
\begin{equation}
\label{eq:simclr}
 -\log\frac{\exp \left(\frac{f_\theta(x)^\intercal  f_\theta(x^+)}{\tau}\right)}
      {
      \sum_{x^- \in \gN_{x}}  \exp\left(\frac{f_\theta(x)^\intercal  f_\theta(x^-)}{\tau}\right) + \exp\left(\frac{f_\theta(x)^\intercal  f_\theta(x^+)}{\tau}\right)
      },
\end{equation}
where $\tau$ denotes a temperature parameter. \rev{More details about representative CL methods are deferred to \msec{sec:diff_cl}.}




\subsection{Backdoor Attacks}

\begin{figure}[!t]
    \centering
    \epsfig{file=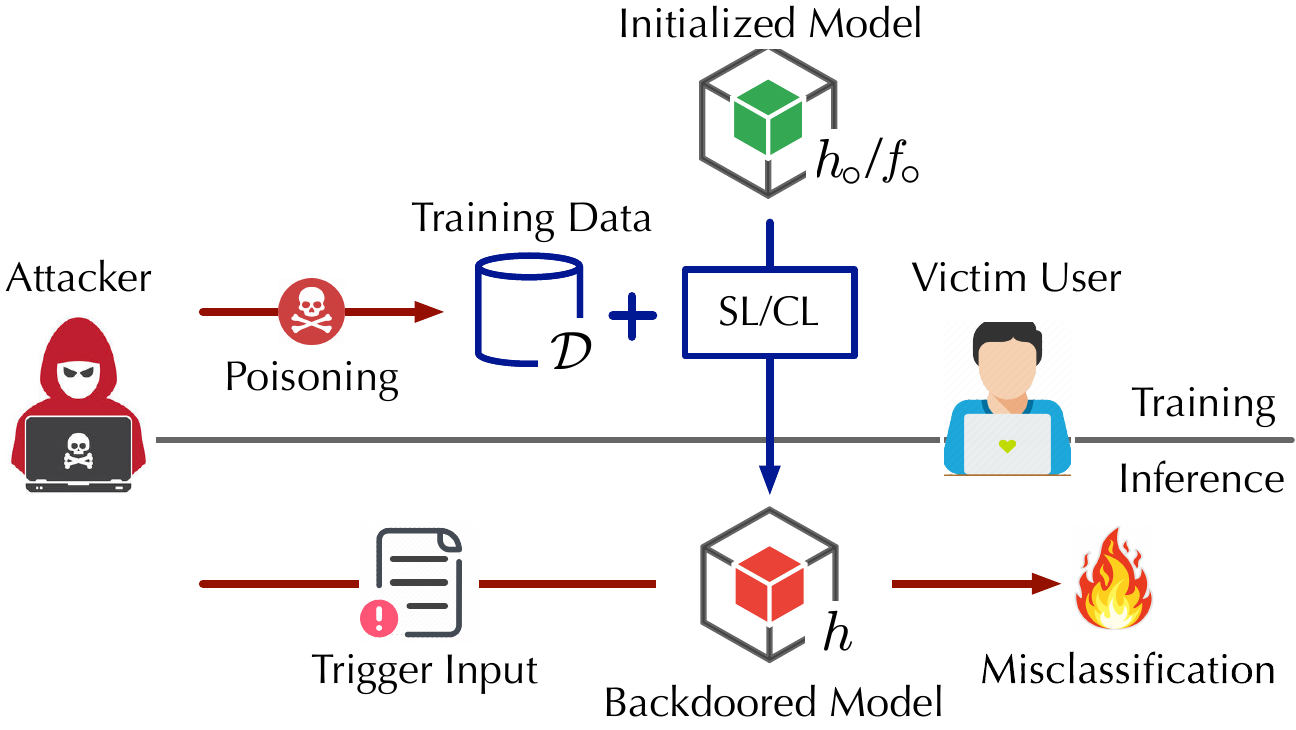, width=75mm}
    \caption{Threat model of \attack.}
    \label{fig:trojan}
\end{figure}

Backdoor attacks represent a major threat to machine learning security\mcite{badnet,trojannn,imc,invisible-backdoor}. As illustrated in Figure\mref{fig:trojan}, the adversary plants a ``backdoor'' into the victim's model during training and activates this backdoor with specific ``triggers'' at inference. The backdoored model reacts to trigger-embedded inputs (trigger inputs) in a highly predictable manner (e.g., classified to the adversary's target class) but functions normally otherwise. Formally, in the supervised setting, the objective function of backdoor attacks via poisoning training data is defined as:
\begin{equation}
\label{eq:trojan_sl}
\min_\theta \sE_{(x, y) \in  \mathcal{D} \cup \mathcal{D^\star}}\, \ell(h_\theta(x), y) ,
\end{equation}
\rev{where $h_\theta$ is the target model (parameterized by $\theta$), $\ell$ denotes the predictive loss (e.g., cross-entropy), and $\gD$ and $\gD^\star$ respectively denote the clean and poisoning training data. Note that $\mathcal{D}^\star$ comprises trigger inputs, which are assigned the target-class label $t$. The poisoning ratio \( |\mathcal{D^\star}| /|\mathcal{D}| \) dictates the influence of clean and poisoned data.}


Backdoor attacks are of particular interest for CL: as CL-trained models are subsequently used in various downstream tasks, backdoor attacks may cause widespread damage\mcite{li2022demystifying}. While supervised backdoor attacks are often inapplicable to CL due to their reliance on labels (cf. \meq{eq:trojan_sl}), recent studies explore new ways of injecting backdoors into CL-trained models\mcite{jia:2021:badencoder,saha:2021:backdoor,liu:2022:poisonedencoder,carlini2021poisoning}.

Specifically, \rev{one class of attacks (e.g., SSLBackdoor\mcite{saha:2021:backdoor}) hypothesizes that during training, the model learns to recognize a trigger that is only present in the target class and has a rigid, slightly variable shape as a key feature. This enables the model to effectively detect the trigger and accurately predict the target class at inference, even in the absence of other semantic features.} The other class of attacks (e.g., CTRL\mcite{li2022demystifying}) uses ``symmetric alignment'': it ensures that the trigger pattern is preserved after augmentations; aligning such augmented makes the trigger pattern appear as semantic features of the target class. Notably, CTRL\mcite{li2022demystifying} achieves attack performance comparable to supervised backdoor attacks, suggesting that CL is also highly vulnerable to backdoor attacks.


\section{A General Attack Framework}
\label{sec:method}

\rev{To study supervised and contrastive backdoor attacks side by side, we first define \attack, a general framework that subsumes a number of supervised and contrastive attacks. We begin by introducing its threat model.}

\subsection{Threat Model}

\rev{Specifically, we focus on data-level backdoor attacks in the vision domain and assume a threat model similar to prior work\cite{badnet,trojannn,imc,saha:2021:backdoor,li2022demystifying}, as illustrated in Figure~\ref{fig:trojan}. }





\vspace{2pt}
\noindent
\textbf{Adversary's Goal:} The adversary aims to inject a backdoor
into a target model $h_\theta$ during training such that at inference time,
$h_\theta$ classifies trigger inputs to an adversary-defined
class $t$ while classifying clean inputs correctly. 
\rev{Formally,  
\begin{equation}
\left\{
\begin{array}{cc} 
h_\theta(x^\star)  = t &  x^\star \text{ is triggered} \\
h_\theta(x)  = y &  x \text{ is clean with ground-truth class } y
\end{array}
\right.
\end{equation}} 

\noindent
\textbf{Adversary's Capability:} \rev{The adversary corrupts the victim's training data with a small amount of poisoning data $\gD^\star$. In the supervised setting, $\gD^\star$ comprises input-class pairs $\{(x^\star, t)\}$, in which $x^\star$ is a trigger input; in the contrastive setting, $\gD^\star$ only comprises trigger inputs $\{x^\star\}$ (without labels). The victim's model $h_\theta$ is then trained on the union of clean data $\gD$ and poisoning data $\gD^\star$ (cf. \meq{eq:trojan_sl}). Note that such attacks are often practical, even if the victim downloads the training data from credible sources\mcite{web-poisoning}.}

\vspace{2pt}
\noindent
\textbf{Adversary's Knowledge:} \rev{We assume a limited knowledge setting:
the adversary has no knowledge about
the concrete model $h_\theta$ or training strategy (e.g., SimCLR) but has access to a small number of target-class inputs sampled from the same distribution as the training data.}

\subsection{Overview of TRL}

\rev{Despite their evident variations, various supervised and contrastive backdoor attacks follow a similar methodology for generating poisoning data, which can be distilled into a high-level framework that we refer to as \attack.} As sketched in Algorithm\mref{alg:ctrl}, \attack involves two key functions to generate poisoning data: $\mathsf{SelectCandidate}()$, which selects candidate inputs from a reference dataset, and $\mathsf{ApplyTrigger}()$, which applies the trigger to each candidate input to generate the poisoning data. 

\begin{algorithm2e}[t]
    \small
    \SetAlgoLined
    \SetKwFunction{ProcedureTitle}{Teste}
    \SetKwProg{KwPcdr}{Procedure}{}{end}
    \DontPrintSemicolon

     \KwIn{$t$: target class; $k$: number of poisoning inputs; $\tilde{\gD}$: reference data}
     \KwOut{$\gD^\star$: poisoning data}
\tcc{\rev{\tt 
description: select candidate inputs from reference data\\
options: (i) from target class only and (ii) across all the classes}}
$\gD^\star \gets \mathsf{SelectCandidate}(\tilde{\gD}, t, k)$;    
    
    \For{$x^\star \in \gD^\star$}
    {
\tcc{\rev{\tt 
description: apply the trigger to an input to generate a trigger input\\
options: (i) universal, (ii) functional, and (iii) dynamic trigger}}    
    update $x^\star \gets \mathsf{ApplyTrigger}(x^\star)$
    }
        
    \KwRet{$\gD^\star$}
    \caption{\attack Attack}
    \label{alg:ctrl}

\end{algorithm2e}

\vspace{2pt}
$\mathsf{SelectCandidate}()$ -- There are typically two ways of selecting candidate inputs, one across all classes while the other exclusively from the target class. 

Among supervised backdoor attacks, dirty-label attacks\mcite{badnet,trojannn,imc} generate poisoning data along with incorrect labels, while clean-label attacks\mcite{shafahi2018poison} do not tamper with the labels of poisoning data. Thus, dirty-label attacks select candidates across all classes, while clean-label attacks select candidates from the target class only. 

Meanwhile, without data labeling, contrastive backdoor attacks rely on aligning positive pairs to associate the trigger with inputs from the target class (details in \msec{sec:comparison}). Therefore, contrastive attacks\mcite{li2022demystifying,saha:2021:backdoor} often select candidates from the target class only.

\vspace{2pt}
$\mathsf{ApplyTrigger}()$ -- There are a variety of trigger definitions, including (i)  universal triggers (e.g., image patch\mcite{badnet,trojannn}), (ii) functional triggers (e.g., specific spectral transformations\mcite{li2022backdoor}), and (iii) dynamic triggers (e.g., input-aware perturbation generated by a generative model\mcite{nguyen2020input}). 

\subsection{Instantiations}
\label{sec:instance}

We now discuss the instantiations of \attack in our studies. 

To make a fair comparison, in both supervised and contrastive settings, we implement $\mathsf{SelectCandidate}()$ as randomly sampling inputs from the target class, corresponding to a clean-label attack.\footnote{Note that as the clean-label variant of the dynamic backdoor attack has poor performance, we use its dirty-label variant, which selects inputs across all classes.}

Further, we instantiate $\mathsf{ApplyTrigger}()$ as follows. A universal trigger is defined as a fixed-size (e.g., 5$\times$5) image patch and applied to a pre-defined position (e.g., left lower corner) of the given input. A functional trigger is defined as a specific spectral transformation to the given candidate input (e.g., increasing the magnitude of a particular frequency by a fixed amount). A dynamic trigger is defined as input-specific perturbation generated by a generative model (e.g., GAN). The detailed implementation of these triggers is deferred to 
\msec{sec:impl}.

Below we use $\sboth{\textsc{Trl}}{\mathrm{SL/CL}}{\mathrm{uni/fun/dyn}}$ to denote a specific backdoor attack, in which SL/CL indicates supervised/contrastive while uni/fun/dyn indicates the trigger type.

\section{Comparison of Supervised and Contrastive Backdoor Attacks}
\label{sec:comparison}

While prior work shows that supervised and contrastive backdoor attacks are comparably effective\mcite{li2022demystifying}, it remains unclear how they differ fundamentally from a defense perspective. Next, we compare supervised and contrastive attacks in terms of learning dynamics and feature distributions.

\subsection{Evaluation Setting}
\label{sec:evalsetting}

We first describe the setting of our evaluation.

{\bf Datasets --}
We primarily use three benchmark datasets: CIFAR10\mcite{cifar}, which contains 60,000 32$\times$32  images divided into 10 classes; CIFAR100\mcite{cifar}, which includes 100 classes, each containing 600 32$\times$32 images; and ImageNet100\mcite{le2015tiny}, which is a subset of the ImageNet-1K dataset with 100 classes, each with 500 training and 50 testing images. 

{\bf CL methods --} We use three representative CL methods \rev{in the vision domain}: SimCLR\mcite{chen:2020:simple}, BYOL\mcite{grill:2020:bootstrap}, and MoCo\mcite{he:2020:momentum}. 

{\bf Attacks --} We consider all the instantiations of \attack attacks in \msec{sec:instance} including both supervised and contrastive attacks as well as different trigger types (i.e., universal, functional, and dynamic). \rev{By default, we allocate 1\% of the training data to construct the poisoning dataset $\gD^\star$ following Algorithm\mref{alg:ctrl} while using the remainder as the clean dataset $\gD$. We consider class 0 as the target class $t$. More implementation details about the attacks are deferred to Table\mref{tab:hyper_attack} in  \msec{sec:setting}.}

{\bf Models --} In both SL and CL, we use ResNet18\mcite{resnet} as the backbone model. In CL, we add a two-layer MLP projector to map the representations to a 128-dimensional latent space. In \msec{sec:discussion}, we also discuss the influence of the backbone model.

{\bf Metrics --} We mainly use two metrics to evaluate the attacks: clean accuracy (ACC) measures the model's accuracy in classifying clean data, and attack success rate (ASR) measures the model's accuracy in classifying poisoning data to the target class.

Table\mref{table:asr_acc} compares the performance of different attacks on benchmark datasets. For contrastive attacks, Table\mref{table:asr_acc} only lists the results of $\sboth{\textsc{Trl}}{\mathrm{CL}}{\mathrm{fun}}$, with other results deferred to \msec{sec:addition}. Note that for a fair comparison, we set the training epochs to ensure that SL and CL attain similar ACCs. The default parameter setting is listed in \msec{sec:setting}.

\begin{table}[t]{\footnotesize
\renewcommand{\arraystretch}{1.2}
 \setlength{\tabcolsep}{1.5pt}
 {
\begin{tabular}{cc|cccccc}
\multicolumn{2}{c|}{\multirow{3}{*}{Dataset}} & \multicolumn{6}{c}{Attack} \\ \cline{3-8} 
\multicolumn{2}{c|}{}                         & $\sboth{\textsc{Trl}}{\mathrm{SL}}{\mathrm{fun}}$  & $\sboth{\textsc{Trl}}{\mathrm{SL}}{\mathrm{uni}}$  & $\sboth{\textsc{Trl}}{\mathrm{SL}}{\mathrm{dyn}}$  & \begin{tabular}[c]{@{}c@{}}\\ SimCLR\end{tabular} & \begin{tabular}[c]{@{}c@{}}$\sboth{\textsc{Trl}}{\mathrm{CL}}{\mathrm{fun}}$\\ BYOL\end{tabular} & \begin{tabular}[c]{@{}c@{}}\\ MoCo\end{tabular} \\ \hline
\multirow{2}{*}{CIFAR10}         & ASR (\%)        &  66.7 & 31.5 & 99.2 & 98.3 & 61.5 & 89.1  \\
                                 & ACC (\%)        &   82.1 & 81.4 & 81.2 & 82.1 & 85.6 & 81.7 \\
     \hline
\multirow{2}{*}{CIFAR100}        & ASR (\%)        &  99.7 & 83.7 & 96.4 &  98.1  & 82.5 & 88.8 \\
                               & ACC (\%)       &   51.4 & 48.3 & 50.1 & 44.2 & 54.5 & 46.7 \\
          \hline
\multirow{2}{*}{ImageNet100}        & ASR (\%)        &  98.4 & 84.2 & 98.2 &  42.4  & 45.1 & 48.1 \\
                               & ACC (\%)    & 52.3    & 52.7 & 55.3 & 47.8 & 49.4 & 44.3
\end{tabular}}
\caption{Clean accuracy and attack success rate of different attacks. \label{table:asr_acc}}}
\end{table}

\subsection{Learning Dynamics}
\label{sec:learn_dyn}

\begin{figure*}[!ht]
\centering
	\includegraphics[width=0.9\linewidth]{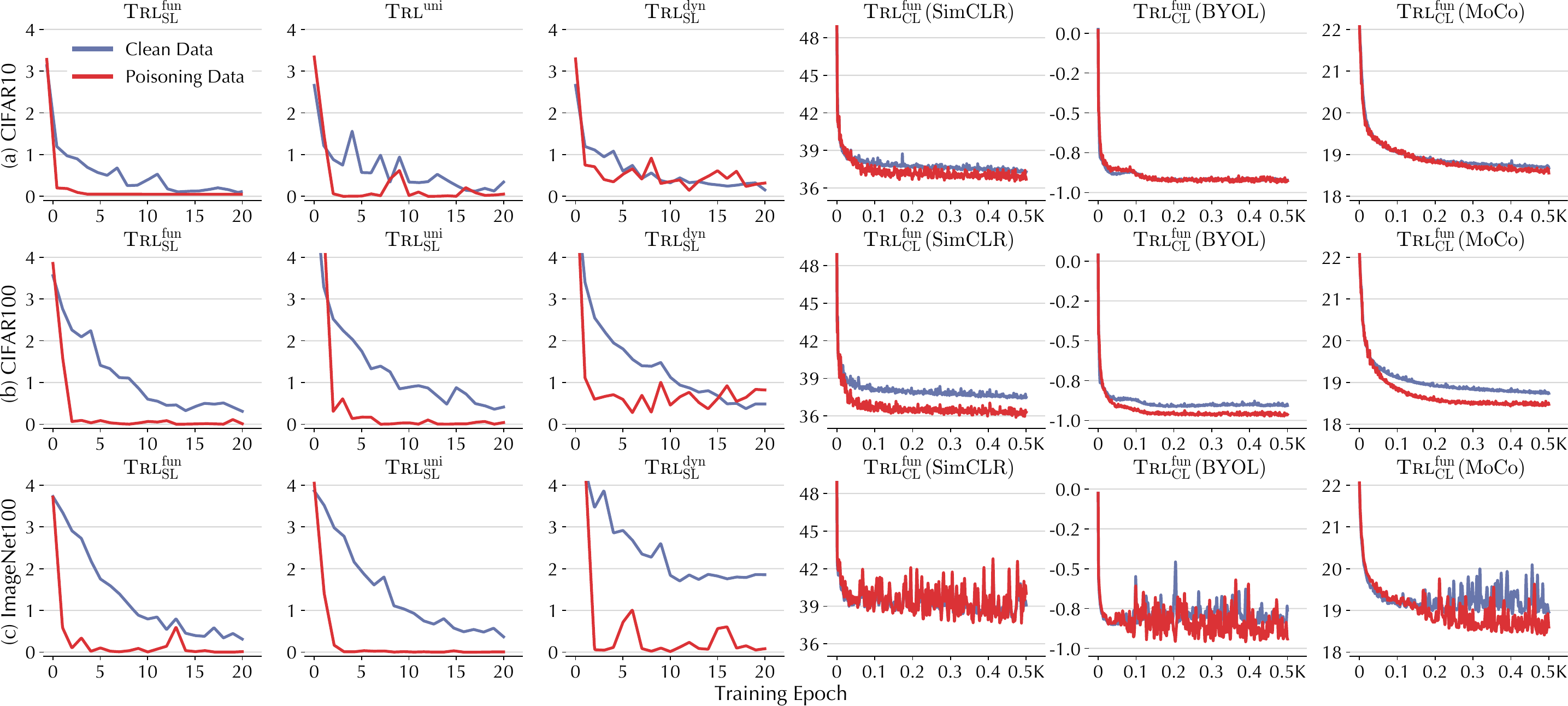}
    \caption{Learning dynamics (measured by the training loss of poisoning and clean data) in supervised and contrastive backdoor attacks on (a) CIFAR10, (b) CIFAR100, and (c) ImageNet100.}
\label{fig:leanring_dynamics}
\end{figure*}

Given training data as input-class pairs ${(x,y)}$, SL optimizes the predictive loss (e.g., measured by cross entropy $H$):
\begin{equation}
\label{eq:sl}
\gL_\mathrm{prd}(\theta) = \sE_{(x, y)} H( h_\theta(x), y),
\end{equation}
where $h$ refers to the end-to-end model $h = g\circ f$. Existing work\mcite{li2021anti,trap-replace} suggests that in supervised attacks, the backdoor task  (i.e., learning the trigger features) is often much ``easier'' than the benign task (i.e., learning the semantic features). Thus, the model often learns the backdoor task much faster than the benign task, reflected in that the predictive loss of poisoning data drops much faster than that of clean data during training.

Due to the absence of labels, CL optimizes the contrastive loss, which measures the similarity between the variants of the same input under different augmentations (positive pairs) with respect to different inputs (negative pairs). For instance, SimCLR\mcite{chen:2020:simple} defines the contrastive loss using InfoNCE loss\mcite{infonce} (cf. \meq{eq:simclr}),
where the numerator term encourages the similarity between positive pair $(x, x^+)$ and the denominator term suppresses the similarity between negative pairs $(x, x^-)$.

To test whether this difference between the learning dynamics of backdoor and benign tasks also holds for contrastive attacks as well, we measure the average contrastive loss of poisoning and clean data during training, with results also summarized in Figure\mref{fig:leanring_dynamics}. 
 

In supervised backdoor attacks, across different trigger settings, the predictive loss of poisoning data decreases faster than that of clean data, which corroborates prior work\mcite{li2021anti}. For example, the loss of poisoning data in $\sboth{\textsc{Trl}}{\mathrm{SL}}{\mathrm{uni}}$ and  $\sboth{\textsc{Trl}}{\mathrm{SL}}{\mathrm{fun}}$ drops sharply to zero in the first few epochs. This difference is especially significant on CIFAR100, which represents a more challenging task in terms of learning semantic features, therefore much more difficult than the backdoor task.

Meanwhile, in contrastive backdoor attacks, across all the CL methods, the training loss of poisoning data decreases gradually at a pace similar to clean data on CIFAR10, CIFAR100, and ImageNet100. While the loss of poisoning data seems slightly lower than clean data on CIFAR100 and ImageNet100 (which may be explained by the more challenging tasks, as indicated by the low clean accuracy in Table\mref{table:asr_acc}), their decreasing rates are highly comparable.

\begin{mtbox}{Remark}
{\small Compared with supervised attacks, in contrastive attacks, the learning dynamics of poisoning and clean data are much less distinguishable.}
\end{mtbox}

\subsection{Feature Distributions}

It is commonly observed in supervised attacks that the poisoning and clean data often form separable clusters in the feature space. This ``latent separability'' premise is exploited by a number of existing defenses against supervised backdoor attacks\mcite{spectral-trigger,chen2018detecting,decoupling,tact,spectre}.

\begin{figure}[t]
\centering
\includegraphics[width=1.0\columnwidth]{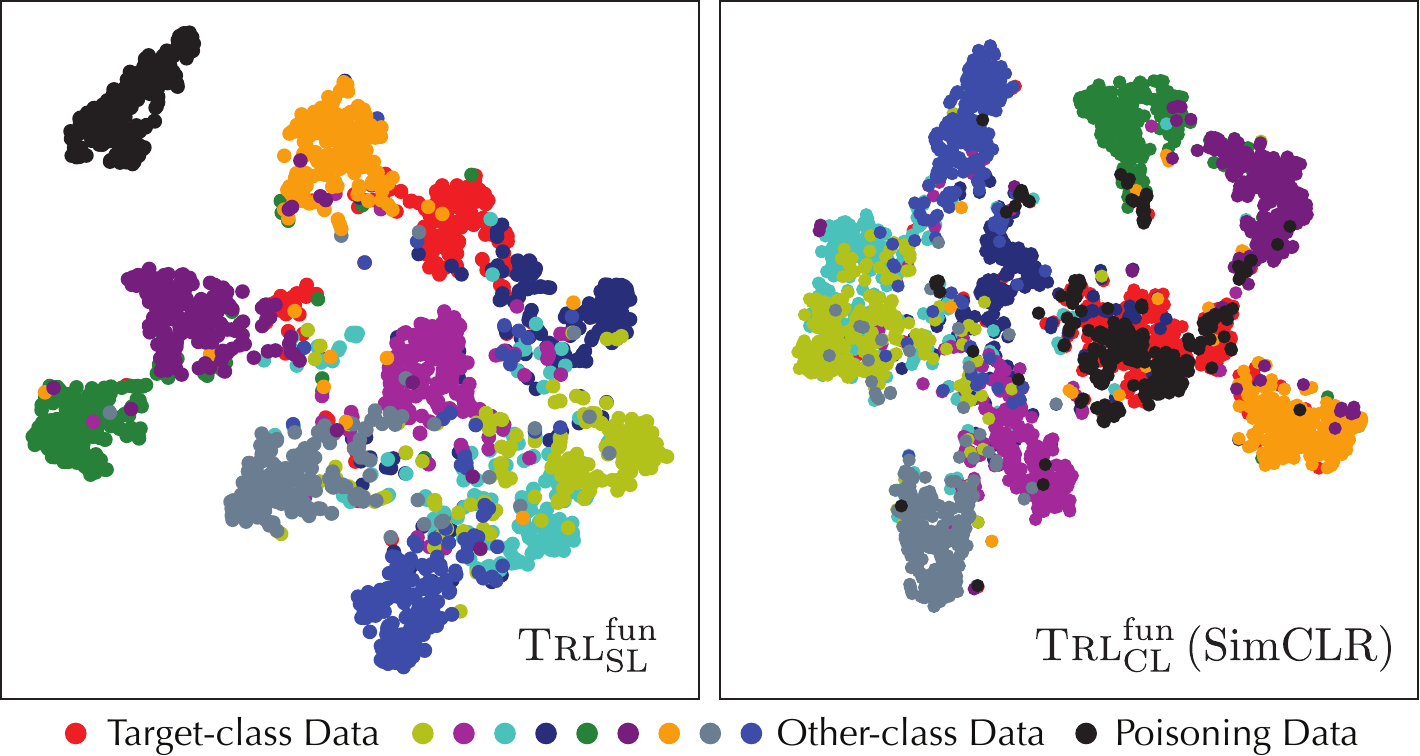} 
\caption{$t$-SNE visualization of the features of clean and poisoning data in $\protect\sboth{\textsc{Trl}}{\mathrm{SL}}{\mathrm{fun}}$ and $\protect\sboth{\textsc{Trl}}{\mathrm{CL}}{\mathrm{fun}}$ on CIFAR10 (target-class data: red; poisoning data: black).}
\label{figure:ssl-feature}
\end{figure}

To test whether this property also holds for contrastive attacks, 
we first use $t$-SNE\mcite{van2008visualizing} to visualize the representations of poisoning and clean data in supervised and contrastive attacks, exemplified by  $\sboth{\textsc{Trl}}{\mathrm{SL}}{\mathrm{fun}}$ and $\sboth{\textsc{Trl}}{\mathrm{CL}}{\mathrm{fun}}$, with results shown in Figure\mref{figure:ssl-feature}. We have the following observations.

\begin{figure*}[!ht]
\centering
\includegraphics[width=0.9\linewidth]{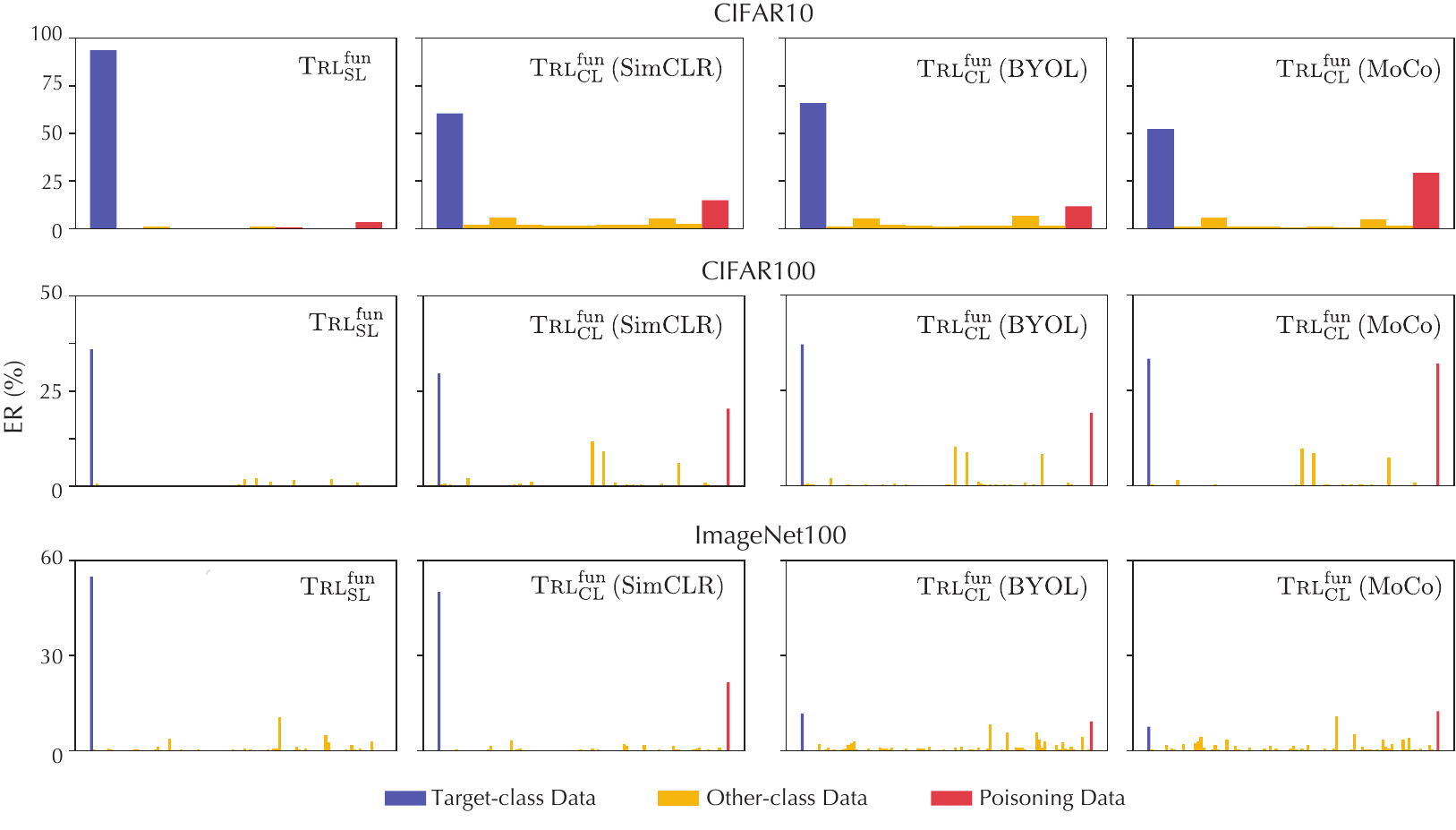}
 
\caption{Entanglement ratio of supervised and contrastive backdoor attacks on CIFAR10, CIFAR100 \rev{and ImageNet100}.}
\label{fig:er_distribution}
\end{figure*}

In supervised attacks, although the clusters of poisoning (in black) and target-class (clean) data (in red) are assigned the same label, they are well separated in the feature space, suggesting that supervised attacks do not necessarily associate the poisoning data with the target-class (clean) data. This finding also corroborates prior work\mcite{tact}. Meanwhile, in contrastive attacks, the clusters of poisoning and target-class data are highly overlapping in the feature space, suggesting that contrastive attacks may take effect by ``entangling'' the representations of poisoning and target-class data.



To further validate this observation quantitatively, we define a metric to measure the ``entanglement effect'' between poisoning and target-class data. Specifically, we define entanglement ratio (ER), which extends the confusion ratio metric\mcite{wang:2022:chaos} to our setting. Specifically,  we randomly sample $n$ inputs per class and $n$ poisoning inputs to form the dataset $\gD$.
In addition, we randomly sample $m$ clean target-class inputs $\gD^t$ (disjoint with $\gD$) as the testing set. For each $x \in \gD^t$, we find its $k$ nearest neighbors $\gN_{k}(f(x))$ among $\gD$ in the feature space and then measure the fraction of such neighbors belonging to each class $c$:
\begin{equation}
\label{eq:metric}
  \mathrm{ER}(f, c) = \frac{1}{k}  \sE_{x \in  \gD^t, x' \in \gD}   \bm{1}_{   h(x') = c \wedge f(x')  \in  \gN_{k}(f(x))}
\end{equation}
where $\bm{1}_p$ is an indicator function that returns 1 if $p$ is true and 0 otherwise and $h(x)$ returns $x$'s class. Consider the poisoning data as a new class $c^\star$. Intuitively, a large  $\mathrm{ER}(f, c^\star)$ suggests that the poisoning and clean target-class inputs are tightly entangled in the feature space.  


We evaluate the entanglement ratio of different attacks with $n = 800$ and $m = 1,000$. \rev{Figure\mref{fig:er_distribution} illustrates the degree of entanglement in the feature space between inputs from each class and that from the target class $t$. Here, each bar represents one distinct class, with the leftmost and rightmost bars corresponding to the target class $t$ and poisoning class $c^\star$, respectively.}
It is observed that the $\mathrm{ER}(f, c^\star)$ of supervised attacks is not significantly different from any other class $c \neq t$. Meanwhile, in contrastive attacks, $\mathrm{ER}(f, c^\star)$ is much higher than the other classes. 
For example, for MoCo on CIFAR100, $\mathrm{ER}(f, c^\star)$ is comparable to $\mathrm{ER}(f, t)$ (around 31\%), indicating that there is tight entanglement between the poisoning and target-class data. 
\begin{mtbox}{Remark}
{\small Compared with supervised attacks, in contrastive backdoor attacks, the poisoning and clean data are highly entangled in the feature space.}
\end{mtbox}

In the evaluation above, we examine the entanglement effect in the feature space (i.e., the output of the last convolutional layer) of the pre-trained encoder. One intriguing question arises: how does this entanglement effect evolve over the model's different layers? To answer this question, we truncate the pre-trained encoder $f$ at a specific ResBlock $k$ at a selected layer $l$, concatenate its part before $(l,k)$ (denoted by $f_{l,k}$) with a classifier $g$ to form an end-to-end model $h_{l,k}$, and fine-tune $g$ using the downstream dataset. 


\begin{table}[!ht]{\footnotesize
\centering
\renewcommand{\arraystretch}{1.1}
\begin{tabular}{cc|cc}
\multicolumn{2}{c|}{Truncated} & \multirow{2}{*}{ACC} & \multirow{2}{*}{ASR} \\
Layer $l$ & Block $k$ & & \\
\hline
\multirow{2}{*}{3}           & 1           & 64.5\%               & 15.2\%               \\
                                  & 2           & 72.1\%               & 54.2\%               \\
                                  \hline
\multirow{2}{*}{4}           & 1           & 79.3\%               & 82.3\%               \\
                                  & 2           & 83.1\%               & 86.4\%              
\end{tabular}
\caption{ACC and ASR of the end-to-end model $h_{l,k}$. \label{table:layers}}}
\end{table}

Table\mref{table:layers} summarizes the performance of the end-to-end model $h_{l,k}$, \rev{with the encoder trained by $\sboth{\textsc{Trl}}{\mathrm{SL}}{\mathrm{fun}}$ using SimCLR}. Observe that $h_{l,k}$'s ACC gradually increases with $(l,k)$. This is expected, as early layers typically extract coarse-grained features, which are then refined throughout subsequent layers. Meanwhile, note that $h_{l,k}$'s ASR also increases with $(l,k)$, suggesting that trigger features and semantic features are not only entangled in the feature space but also intertwined along the feature extraction process. 

\begin{figure}[!ht]
	\centering
	\includegraphics[width=0.9\columnwidth]{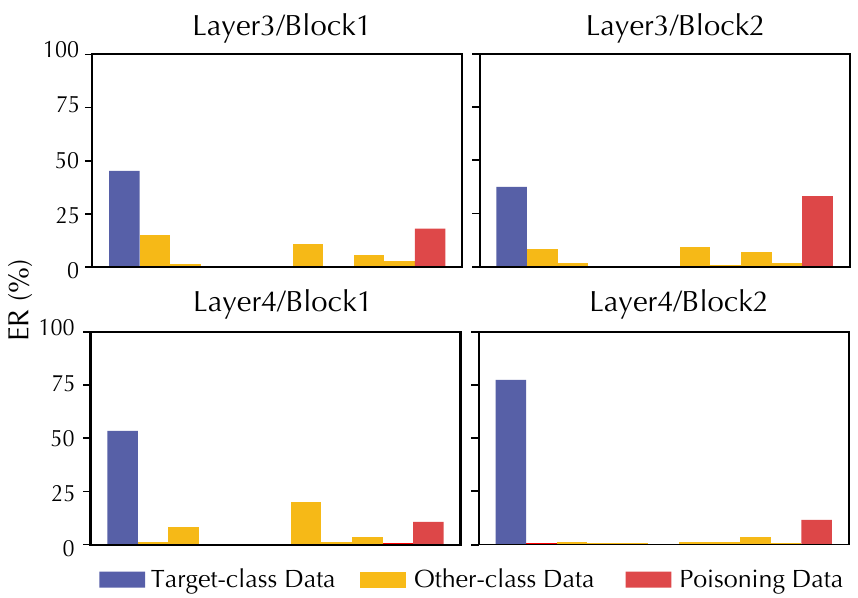}
	\caption{Entanglement ratios of the truncated encoder $f_{l,k}$. \label{fig:er_layers}}
\end{figure}

We further examine the entanglement ratios (cf. \meq{eq:metric}) of the truncated encoder $f_{l,k}$ for different $(k,l)$, with results summarized in Figure\mref{fig:er_layers}, \rev{following the same setting in Figure\mref{fig:er_distribution}.
Observe that despite the apparent variations of different truncated encoders,} the entanglement ratios of different $(k,l)$ show patterns similar to Figure\mref{fig:er_distribution}, suggesting that the entanglement effect may appear across different layers of the encoder. 
\begin{mtbox}{Remark}
{\small In contrastive backdoor attacks, trigger and semantic features are not only entangled in the feature space but also intertwined throughout the feature extraction process.}
\end{mtbox}

\begin{figure*}[!ht]
	\centering
	\includegraphics[width=0.8\linewidth]{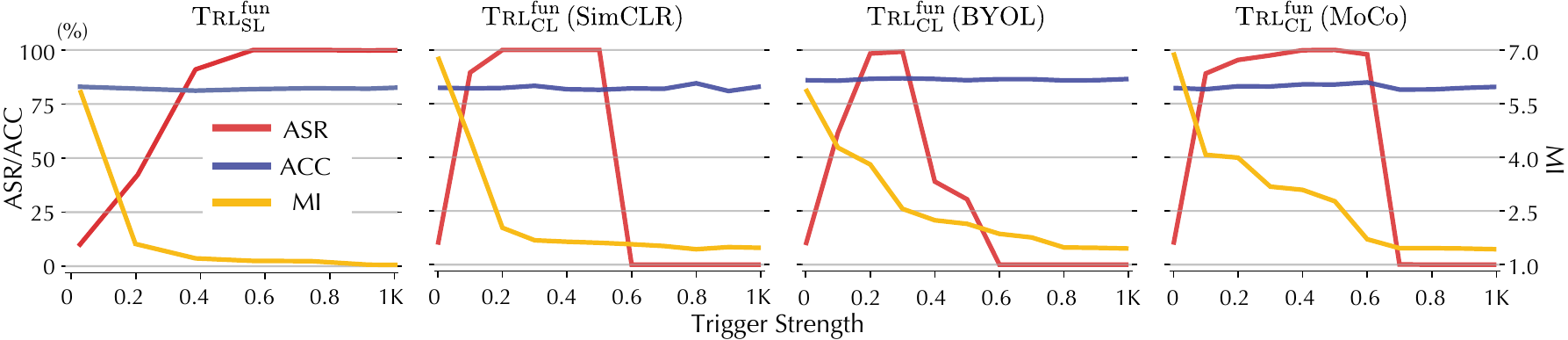}
	\caption{ASR, ACC, and MI of different attacks with respect to the trigger strength (measured by the magnitude of spectral perturbation). \label{fig:overlap2}}
\end{figure*}

\section{Possible Explanations}
\label{sec:explanation}

The empirical study shows that supervised and contrastive backdoor attacks demonstrate distinctive learning dynamics and feature distributions. Below we provide possible explanations for such phenomena, revealing that the two classes of attacks operate through different mechanisms.

\subsection{Preliminaries}

We begin by introducing two key concepts.

\vspace{2pt}
{\bf Trigger strength} (TS) quantifies the weight of trigger features in poisoning inputs. Specifically, in our setting, we define the trigger strength of $\sboth{\textsc{Trl}}{\mathrm{SL/CL}}{\mathrm{dyn/uni}}$ as the size of the trigger image patch (e.g., 5$\times$5), under a fixed transparency setting.
We define the trigger strength of $\sboth{\textsc{Trl}}{\mathrm{SL/CL}}{\mathrm{fun}}$ as the perturbation magnitude in the spectral domain (e.g., increasing the magnitude of a specific frequency band by 50).
Intuitively, by adjusting the trigger strength, we balance the importance of trigger features and semantic features in poisoning inputs.

\vspace{2pt}
{\bf Mutual information} (MI) quantifies the amount of information carried by one random variable about another. In our context, we employ MI to estimate the proportion of semantic features retained in the representations of poisoning inputs compared to their corresponding clean inputs. Specifically, \rev{consider the representations (i.e., the encoder's outputs) of a clean input and its poisoning counterpart as two random variables $X$ and $X'$}, we estimate their MI $I(X; X')$ following\mcite{kraskov2004estimating}.
Specifically, let $z_i = (x_i, x_i')$ be a point sampled from $(X, X')$. We define the distance between $z_i$ and $z_j$ as 
\rev{\begin{equation}
    \| z_i -z_j \| = \max\{ \|x_i - x_j \|_2, \| x_i' - x_j'\|_2 \}
\end{equation}}
Let $\epsilon(i)/2$ be the distance from $z_i$ to its $k$-th nearest neighbor among the given set of points and $\epsilon_x(i)/2$ and $\epsilon_{x'}(i)/2$ be the same distance projected to the $X$ and $X'$ subspaces. Further, let $n_x(i)$ (or $n_{x'}(i)$ be the number of points with distance to $x_i$ (or $x_i'$) less than $\epsilon_x(i)/2$ (or $\epsilon_{x'}(i)/2$). Then, the MI of $X$ and $X'$ is quantified by:
\begin{equation}
\label{eq:metric3}
I(X;X') = \psi(k)  - \frac{1}{k} - \langle \psi(n_{x}) + \psi(n_{x'}) \rangle + \psi(N)
\end{equation}
where $\psi$ is the digamma function, $k$ is a hyper-parameter, and $N$ is the number of points in $X$ (or $X'$). We use \meq{eq:metric3} to measure the proportion of semantic features retained in the representations of poisoning inputs compared to their corresponding clean inputs. In the implementation, we set $N=2,000$ and $k=5$.

\subsection{Supervised Backdoor Attacks}

Under the supervised setting, the backdoor attack is implemented as optimizing the following objective function:
\begin{equation}
\label{eq:trojan_sl2}
\min_\theta \gL_\mathrm{prd}(\theta) + \lambda  \gL^\star_\mathrm{prd}(\theta) 
\end{equation}
where $\gL$ and $\gL^\star$ denote the loss function defined in \meq{eq:sl}, measured on clean data $\gD$ and poisoning data $\gD^\star$ respectively, and the hyper-parameter $\lambda$ balances the influence of $\gD$ and $\gD^\star$ (e.g., via controlling $|\gD^\star|/|\gD|$). Intuitively, the first and second terms represent the benign task (i.e., learning semantic features) and backdoor task (i.e., learning trigger features), respectively. We hypothesize that in supervised attacks, if the trigger feature is sufficiently evident, the learning of benign and backdoor tasks tends to occur independently at the data level; that is, it focuses on learning the backdoor task from the poisoning data and focuses on learning the benign task from the clean data. 

To validate this hypothesis, we assess the effect of trigger strength on the ASR, ACC, and MI of supervised backdoor attacks, with results shown in Figure\mref{fig:overlap2}. Observe that as the trigger strength varies from 0 to 1K, the ASR of $\sboth{\textsc{Trl}}{\mathrm{SL}}{\mathrm{fun}}$ increases rapidly and remains around 100\%, while its MI quickly drops and stays at the minimum level. This observation suggests that in successful attacks, the model focuses on learning trigger features (backdoor task) only from the poisoning data. Meanwhile, the trigger strength has little impact on the ACC, indicating that the model effectively learns the benign task from the clean data.
Similar trends are also observed on other supervised attacks in \msec{sec:factor}. 
\begin{mtbox}{Remark}
{\small In supervised backdoor attacks, the learning of benign and backdoor tasks occurs independently at the data level, with a focus on learning trigger features from poisoning data and semantic features from clean data.}
\end{mtbox}

The independence of benign and backdoor tasks in supervised attacks explains the empirical observations in \msec{sec:comparison}. In terms of learning dynamics, as the trigger feature (e.g., a specific image patch) is often simpler than the varying semantic features, the model learns the backdoor task much faster than the benign task. 
In terms of feature distributions, as \meq{eq:trojan_sl2} does not specify any constraints on the latent representations of poisoning and clean data, although associated with the same class, the representations of poisoning and target-class data are not necessarily proximate in the feature space.

\subsection{Contrastive Backdoor Attacks}

Meanwhile, under the contrastive setting, the backdoor attack is implemented as the following objective function:
\begin{equation}
\label{eq:trojan_cl}
\min_\theta \gL_\mathrm{ctr}(\theta) + \lambda  \gL^\star_\mathrm{ctr}(\theta) 
\end{equation}
where $\gL_\mathrm{ctr}$ and $\gL_\mathrm{ctr}^\star$ are the contrastive loss defined in \meq{eq:simclr}, measured on $\gD$ and $\gD_\star$ respectively. Because there are no labels, contrastive attacks manipulate the distributions of poisoning and target-class data in the feature space. To do so, the model must learn both trigger and semantic features from the poisoning data, so that the semantic features intertwine poisoning data with target-class data, while the trigger features connect all poisoning data. 

\begin{figure}[!ht]
	\centering
	\includegraphics[width=0.6\columnwidth]{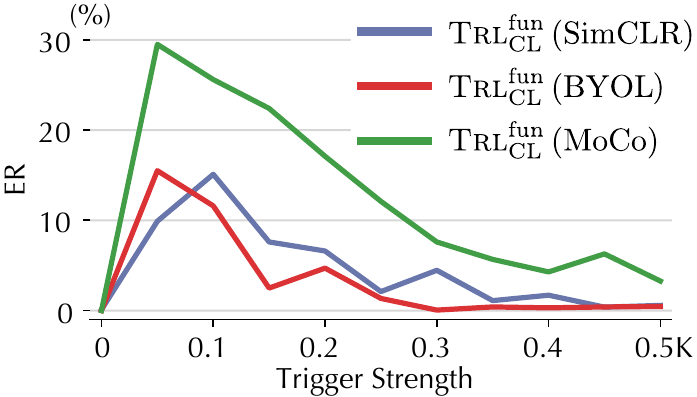}
	\caption{Entanglement effect with respect to trigger strength. \label{fig:overlap1}}
\end{figure}

To confirm our analysis, we assess the impact of trigger strength on the ASR, ACC, and MI of contrastive backdoor attacks, with results shown in Figure\mref{fig:overlap2}. Across all the CL methods, as the trigger strength varies from 0 to 1K, the ASR of $\sboth{\textsc{Trl}}{\mathrm{CL}}{\mathrm{fun}}$ first increases to 100\% and then drops quickly, while its MI gradually decreases during the process. The finding suggests that the optimal attack is attained only if the semantic features are partially retained in the representations of poisoning data. This can be explained by the fact that if the trigger features are insignificant, the semantic features dominate the poisoning data, resulting in weak connections among poisoning data; meanwhile, if the trigger features dominate the poisoning data, the weak semantic features negatively impact the entanglement between poisoning and target-class data. To validate this explanation, we further measure the entanglement between poisoning and target-class data under varying trigger strength. As shown in Figure\mref{fig:overlap1}, the entanglement ratio is not a monotonic function of the trigger strength but follows a similar trend as the ASRs in Figure\mref{fig:overlap2}.
We can thus conclude:
\begin{mtbox}{Remark}
{\small In contrastive backdoor attacks, the learning of benign and backdoor tasks is intertwined in the poisoning data.}
\end{mtbox}

This interdependence between benign and backdoor tasks in contrastive backdoor attacks easily explains the empirical observations in \msec{sec:comparison}. In terms of learning dynamics, because the model learns both benign and backdoor tasks (the latter is simpler than the former) from poisoning data, the contrastive loss of poisoning data tends to decrease at a rate similar to clean data. In terms of feature distributions, as the model extracts both trigger and semantic features from poisoning data, it naturally intertwines poisoning and target-class data in the feature space.

\section{Defense Implications}
\label{sec:defense}

We now explore the implications of the unique characteristics of contrastive attacks from a defense perspective.  

\subsection{Learning Dynamics-Based Defenses}

We first examine defenses premised on the difference between the learning dynamics of benign and backdoor tasks.

\vspace{2pt}
{\bf Anti-Backdoor Learning} (ABL)\mcite{li2021anti} is a filtering defense that segregates poisoning data and unlearns backdoored models. Intuitively, as the backdoor task is simpler than the benign task, assuming the loss of poisoning data drops quickly in early training epochs while the loss of clean data decreases at a steady pace, ABL detects poisoning data and unlearns the backdoored model by applying gradient descent on the clean data and gradient ascent on the detected poisoning data.

Despite its effectiveness against supervised attacks, it is challenging to extend ABL to CL because the learning dynamics of poisoning data in contrastive attacks are much less distinctive (cf.\,\msec{sec:learn_dyn}). We empirically validate this by applying ABL in detecting the poisoning data of $\sboth{\textsc{Trl}}{\mathrm{SL}}{\mathrm{fun}}$ and $\sboth{\textsc{Trl}}{\mathrm{CL}}{\mathrm{fun}}$ under the same setting as in \msec{sec:evalsetting}. 
We follow the hyper-parameter setting in\mcite{li2021anti}. \rev{Specifically, if the loss of any training input goes below a threshold $\gamma =0.5$, we activate gradient ascent to boost its loss to $\gamma$.} In the CL setting, we set $\gamma$ according to a fixed ratio compared to the largest loss drop: $(\gamma - \ell_\text{final}) / (\ell_\text{init} - \ell_\text{final})$, where $\ell_\text{init}$ and $\ell_\text{final}$ respectively denote the loss of the training input before and after training. \rev{Table\mref{table:abl} reports ABL's performance under the  poisoning rate fixed as 1\%. The false positive rate (FPR) measures the fraction of clean inputs falsely detected as poisoning, the true positive rate (TPR) measures the fraction of poisoning inputs correctly identified, and the isolation rate denotes the percentage of samples isolated by ABL. A lower FPR or a higher TPR indicates more accurate detection.}

\begin{table}[!ht]{\footnotesize
\centering
\renewcommand{\arraystretch}{1.2}
\begin{tabular}{c|cc|cc}
\multirow{2}{*}{Isolation Rate} & \multicolumn{2}{c|}{$\sboth{\textsc{Trl}}{\mathrm{CL}}{\mathrm{fun}}$} & \multicolumn{2}{c}{$\sboth{\textsc{Trl}}{\mathrm{SL}}{\mathrm{fun}}$} \\ \cline{2-5} 
                                & TPR            & FPR            & TPR            & FPR           \\ \hline
1\%                             & 2.4\%          &  0.99\%   &       92.8\%         &    0.07\%           \\
5\%                             & 39.6\%         &   4.8\%      &     99.4\%           &  4.1\%             \\
10\%                            & 56.0\%         &   9.5\%     &      99.8\%          &    9.1\%           \\
20\%                            &  63.6\%        &  19.6\%        &   99.8\%             &      19.2\%        
\end{tabular}
\caption{ABL against $\protect\sboth{\textsc{Trl}}{\mathrm{SL}}{\mathrm{fun}}$ and $\protect\sboth{\textsc{Trl}}{\mathrm{CL}}{\mathrm{fun}}$ on CIFAR10. \label{table:abl}}}
\end{table}

Observe that ABL effectively detects the poisoning data of $\sboth{\textsc{Trl}}{\mathrm{SL}}{\mathrm{fun}}$ but is much less effective against $\sboth{\textsc{Trl}}{\mathrm{CL}}{\mathrm{fun}}$. 
For example, with 1\% isolation rate, ABL detects 92.8\% of the poisoning data of $\sboth{\textsc{Trl}}{\mathrm{SL}}{\mathrm{fun}}$, which drops to 2.4\% against $\sboth{\textsc{Trl}}{\mathrm{CL}}{\mathrm{fun}}$. The increase of TPR against $\sboth{\textsc{Trl}}{\mathrm{CL}}{\mathrm{fun}}$ with the isolation rate is mainly attributed to the low poisoning rate (1\%). We thus conclude that ABL is ineffective against contrastive attacks, which confirms the observations in \msec{sec:learn_dyn}. This is due to the intertwined learning dynamics of both benign and backdoor tasks.  \rev{We defer the evaluation of ABL against $\protect\sboth{\textsc{Trl}}{\mathrm{SL/CL}}{\mathrm{uni}}$ to \msec{sec:app_abl}.}

\subsection{Feature Distribution-Based Defenses}
\label{sec:feature-based-defense}

The entanglement between the representations of poisoning and clean data also causes challenges for defenses that rely on the separability of poisoning data in the feature space.

\begin{figure}[!ht]
	\centering
	\includegraphics[width=0.8\columnwidth]{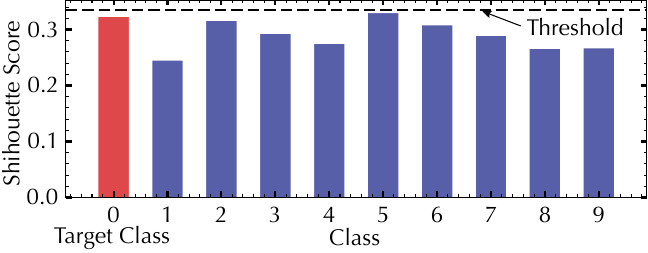}
	\caption{AC against $\protect\sboth{\textsc{Trl}}{\mathrm{CL}}{\mathrm{fun}}$ (SimCLR) on CIFAR10. \label{fig:ac}}
\end{figure}

\begin{figure*}[!ht]
	\centering
	\includegraphics[width=0.85\linewidth]{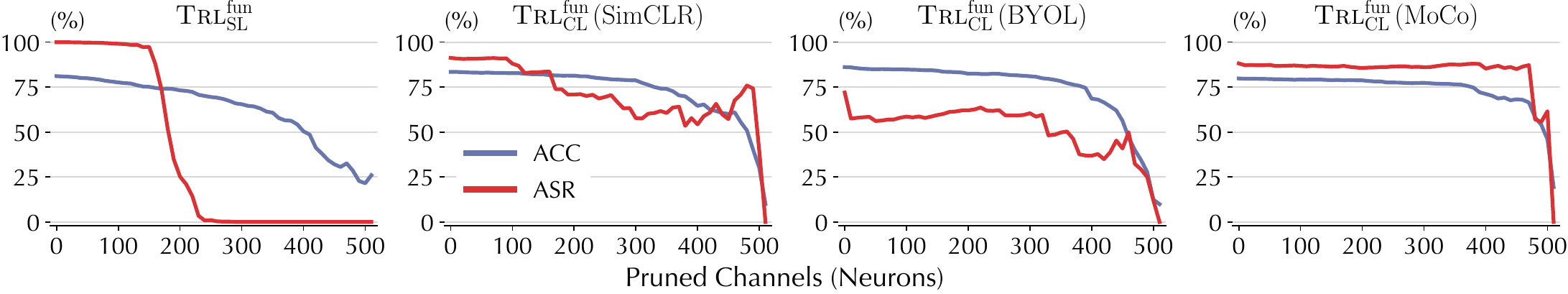}
	\caption{FP against $\protect\sboth{\textsc{Trl}}{\mathrm{CL}}{\mathrm{fun}}$ and  $\protect\sboth{\textsc{Trl}}{\mathrm{SL}}{\mathrm{fun}}$ on CIFAR10. \label{fig:fp}}
\end{figure*}

\vspace{2pt}
{\bf Activation clustering} (AC)\mcite{chen2018detecting} is a data inspection method. It trains the model using the potentially poisoning data and collects the penultimate-layer activation of each input. 
AC assumes the poisoning data in the target class forms a separate cluster that is either small or far from the class center. It identifies the target class by calculating the silhouette score of each class, with a higher score indicating a stronger fit to two clusters. Additionally, since the attacker is assumed to be unable to poison more than half of the data, the smaller cluster is considered to be the poisoning data. Assuming the data labels are available, we evaluate the effectiveness of AC against both supervised and contrastive attacks. We set the hyper-parameters according to the original paper.

Figure\mref{fig:ac} shows that AC fails to identify the target class (class 0), which has a lower score compared to other classes (e.g., class 5), not to mention detecting the poisoning data. This may be attributed to the tight entanglement between the representations of poisoning and clean data.

\vspace{2pt}
{\bf Statistical contamination analyzer} (SCAn)\mcite{tact} detects poisoning data based on the statistical properties of its representations. Specifically, it applies an EM algorithm to decompose an input into two terms: identity and variance; it then analyzes the representations in each class and identifies the target class most likely to be characterized by a mixture model. To evaluate SCAn against contrastive attacks, following\mcite{tact}, we use 1,000 inputs randomly sampled from the testing set to build the decomposition model, which we apply to analyze 5,000 poisoning and 5,000 clean inputs. For comparison, we also evaluate SCAn against the supervised attack $\sboth{\textsc{Trl}}{\mathrm{SL}}{\mathrm{fun/uni}}$. Similar to STRIP\mcite{gao2019strip}, we vary the FPR under three different FPR settings: 0.5\%, 1.0\%, and 2.0\%,  and evaluate the TPR.

\begin{table}[t]{\footnotesize
\centering
\renewcommand{\arraystretch}{1.2}
\begin{tabular}{c|cccc}
\multirow{2}{*}{FPR} & \multicolumn{4}{c}{TPR}    \\
\cline{2-5}
                     & $\sboth{\textsc{Trl}}{\mathrm{CL}}{\mathrm{fun}}$  & $\sboth{\textsc{Trl}}{\mathrm{SL}}{\mathrm{fun}}$  & $\sboth{\textsc{Trl}}{\mathrm{CL}}{\mathrm{uni}}$ & $\sboth{\textsc{Trl}}{\mathrm{SL}}{\mathrm{uni}}$\\ \hline
0.5\%                    & 28.0\%    & 63.0\%    &  28.4\%  &   51.5\% \\
1.0\%                & 28.0\%    & 66.5\%         &  63.3\%  &   71.8\%      \\
2.0\%                  & 28.0\%    & 68.0\%       &  76.7\%  &    85.5\%    \\ 
\end{tabular}
\caption{SCAn against $\protect\sboth{\textsc{Trl}}{\mathrm{CL}}{\mathrm{fun}}$ (SimCLR) and $\protect\sboth{\textsc{Trl}}{\mathrm{SL}}{\mathrm{fun}}$on CIFAR10. \label{table:scan}}}
\end{table}

\rev{As shown in Table\mref{table:scan}, compared with the supervised attack, SCAn is much less effective, especially for $\sboth{\textsc{Trl}}{\mathrm{CL}}{\mathrm{fun}}$.} For instance, SCAn detects over 63.0\% of the poisoning data of $\sboth{\textsc{Trl}}{\mathrm{SL}}{\mathrm{fun}}$, which drops to  28.0\% against $\sboth{\textsc{Trl}}{\mathrm{CL}}{\mathrm{fun}}$. This can be attributed to the entanglement between the representations of poisoning and clean data, while SCAn relies on the separability of poisoning data in the feature space.

\vspace{2pt}
{\bf Fine-pruning} (FP)\mcite{liu2018fine} assumes poisoning and clean data activate different neurons and sanitizes backdoored models by pruning neurons that are dormant with respect to clean data. Specifically, it measures the average activation of each neuron with respect to clean data from a validation set and prunes the set of least active neurons. Following\mcite{liu2018fine}, we apply FP on the penultimate layer of the backdoored model. In\mcite{liu2018fine}, it keeps pruning the model until the tolerance of accuracy (5\%) reduction is reached. For a more detailed analysis, we show the ASR and ACC with respect to the variation of the pruned channels from 0 to 500.

Figure\mref{fig:fp} shows the ASR and ACC of each attack as a function of the pruning rate of FP. Observe that FP effectively defends against supervised attacks. By pruning about 240 neurons, the ASR of  $\sboth{\textsc{Trl}}{\mathrm{SL}}{\mathrm{fun}}$ is reduced to zero while incurring an acceptable ACC drop. Meanwhile, the ASR and ACC of contrastive attacks decrease at a similar pace. To attain effective defenses, it is necessary to prune almost all the neurons. This further suggests the tight entanglement between poisoning and clean data in the feature space.
\begin{mtbox}{Remark}
{\small The unique characteristics of contrastive backdoor attacks render defenses based on learning dynamics and feature distributions ineffective.}
\end{mtbox}

\subsection{Downstream Defenses}

Thus far, we have demonstrated the challenges of defending against contrastive backdoor attacks based on either learning dynamics or feature distributions. Next, we consider a more diverse set of defenses applied to the end-to-end model $h = g \circ f$ that comprises the backdoored encoder $f$ and a classifier $g$ in the downstream task.

\vspace{2pt}
{\bf NeuralCleanse} (NC)\mcite{Wang:2019:sp} is a model inspection method to detect backdoors. \rev{Specifically, for each class, NC reverse-engineers the ``minimal'' trigger (measured by the $L_1$-norm of its binary mask) required to misclassify clean inputs from all other classes into this class. It then runs outlier detection and marks any class with a trigger significantly smaller than the other classes as an outlier and infected. We evaluate NC's efficacy in both supervised and contrastive learning. In the contrastive setting, an end-to-end model, in which the encoder trained by $\sboth{\textsc{Trl}}{\mathrm{CL}}{\mathrm{fun}}$ or $\sboth{\textsc{Trl}}{\mathrm{CL}}{\mathrm{uni}}$ using SimCLR, is fine-tuned using clean data. In the supervised setting, the model is directly trained by $\sboth{\textsc{Trl}}{\mathrm{SL}}{\mathrm{fun}}$ or $\sboth{\textsc{Trl}}{\mathrm{SL}}{\mathrm{uni}}$.}

\begin{table}[t]{\footnotesize\rev{
\centering
\renewcommand{\arraystretch}{1.2}
\setlength{\tabcolsep}{4pt}
\begin{tabular}{c|cc|cc}
        &  $\protect\sboth{\textsc{Trl}}{\mathrm{CL}}{\mathrm{fun}}$ &  $\protect\sboth{\textsc{Trl}}{\mathrm{SL}}{\mathrm{fun}}$ &  $\protect\sboth{\textsc{Trl}}{\mathrm{CL}}{\mathrm{uni}}$ &  $\protect\sboth{\textsc{Trl}}{\mathrm{SL}}{\mathrm{uni}}$ \\ \hline
Anomaly Index & 0.72 $\pm$ 0.38 & 1.47 $\pm$ 0.54 & 1.09 $\pm$ 0.49 & 4.57 $\pm$ 1.28
\end{tabular}
\caption{Effectiveness of NC on CIFAR10. \label{table:nc}}}}
\end{table}

As shown in Table\mref{table:nc}, NC failed to detect any backdoors in the contrastive setting. For example, for $\protect\sboth{\textsc{Trl}}{\mathrm{CL}}{\mathrm{fun}}$, the anomaly index of the target class varies around 0.72, much lower than the detection threshold of 2. This is attributed to the presence of functional triggers and aggressive data augmentation methods, which make it challenging for NC to reverse-engineer a stable trigger. \rev{In contrast, NC successfully detects backdoors against supervised attacks, especially $\protect\sboth{\textsc{Trl}}{\mathrm{SL}}{\mathrm{uni}}$.} 


\vspace{2pt}
{\bf STRIP}\mcite{gao2019strip} is an inference-time data inspection method. Specifically, it assumes that the prediction of poisoning data is insensitive to perturbation due to the dominance of trigger features, whereas the prediction of clean data may vary greatly with perturbation. For an incoming input, STRIP intentionally perturbs it (e.g., superimposing various image patterns) and utilizes the randomness of its prediction to determine whether it is triggered. As it is a run-time detection method, we use an end-to-end model (consisting of the backdoored encoder and downstream classifier) to evaluate its effectiveness. \rev{Following\mcite{gao2019strip}, we first estimate the entropy distribution of clean inputs on a validation set, select an entropy detection threshold with a given FPR (0.5\%, 1.0\%, and 2.0\%), and remove all training inputs with entropy below this threshold.}

\begin{table}[t]{\footnotesize
\centering
\renewcommand{\arraystretch}{1.2}
\begin{tabular}{c|c|c}
 Detection Threshold & FPR   & TPR \\ \hline
      0.56  & 0.5\%                                            &  0.5\%   \\
            0.65       & 1.0\%                               &  1.1\%   \\
            0.75 & 2.0\%                                    &   2.0\% 
\end{tabular}
\caption{STRIP against $\protect\sboth{\textsc{Trl}}{\mathrm{CL}}{\mathrm{fun}}$ (SimCLR) on CIFAR10. \label{table:strip}}}
\end{table}

As shown in Table\mref{table:strip}, STRIP is ineffective in detecting the poisoning data of $\sboth{\textsc{Trl}}{\mathrm{CL}}{\mathrm{fun}}$. For instance, with FRR fixed as 1.0\%, the TPR of STRIP is as low as 1.1\%, suggesting that STRIP fails to distinguish poisoning and clean data. This may also be explained by the entanglement effect: as the poisoning and clean data share similar representations, the perturbation to them leads to similar measures.  \rev{The results on $\sboth{\textsc{Trl}}{\mathrm{CL/SL}}{\mathrm{uni}}$ are deferred to \msec{sec:app_strip}.}

\vspace{2pt}
{\bf Backdoor defense via decoupling} (BDD) \cite{huang2021backdoor} is designed to mitigate the impact of poisoning inputs that tend to cluster together in the feature space of compromised DNNs. This approach consists of three stages: (1) the backbone is trained in a self-supervised manner, which aligns inputs with the same ground-truth label in the feature space; (2) the remaining classifier is trained in a standard manner, using all labeled training inputs; and (3) low-confidence inputs are removed based on the end-to-end  model.

\begin{figure}[t]
	\centering
	\includegraphics[width=0.55\columnwidth]{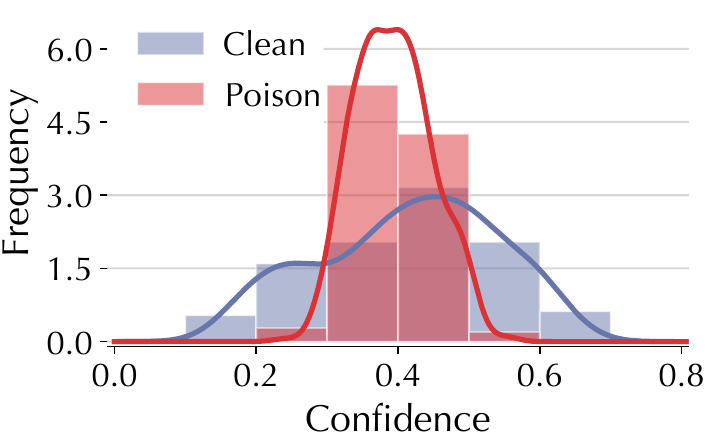}
	\caption{Distributions of the prediction confidence of poisoning and clean data. \label{fig:confidence}}
\end{figure}

Clearly, (1) and (2) share the same settings as the threat model of \attack. Thus, we focus on (3) to evaluate the effectiveness of BDD. The distributions of the classification confidence of poisoning and clean (from the target class) inputs are shown in Figure\mref{fig:confidence}. Note that poisoning inputs not only have similar classification confidence as clean inputs but also exhibit lower variance. This makes it more challenging to establish a confidence threshold to distinguish between them. Consequently, using BDD mechanisms to defend against \attack tends to be ineffective.

\vspace{2pt}
{\bf Fine-tuning}\mcite{liu2017neural} is a method to adapt a pre-trained encoder to a specific downstream task by fine-tuning a classifier. Prior work\mcite{liu2017neural} shows that fine-tuning may mitigate backdoor attacks to a certain extent. In the previous evaluation, following\mcite{saha:2021:backdoor}, we focus on scenarios where the downstream and pre-training datasets are identical. Here, we explore cases in which these datasets differ, such as pre-training the encoder on CIFAR10 and fine-tuning the classifier on CIFAR100. Since the datasets may have varied class structures, we focus on untargeted attacks and measure the attack effectiveness by the model's accuracy drop in classifying trigger inputs compared to clean inputs.

Table\mref{table:finetune} summarizes the results. Observe that the contrastive backdoor attack substantially affects the model's performance, regardless of the setting of pre-training and downstream datasets. For instance, when the encoder is pre-trained on CIFAR100 using SimCLR and then adapted to CIFAR10, the model achieves 53.3\% accuracy and 18.1\% accuracy (i.e., 35.2\% accuracy drop) on clean and trigger inputs, respectively. This observation indicates that fine-tuning using downstream datasets may not mitigate contrastive backdoor attacks. Additionally, we observe that most trigger inputs are misclassified into a single incorrect class. We speculate that these trigger inputs are clustered together and located close to a specific class in the feature space, which is also close to the target class in the pre-training dataset.

\begin{table}[!ht]{\footnotesize
\centering
\renewcommand{\arraystretch}{1.2}
\begin{tabular}{cl|ccc}
\multicolumn{2}{c|}{\multirow{2}{*}{\begin{tabular}[c]{@{}c@{}}Pre-training / Downstream \\ Dataset\end{tabular}}} & \multirow{2}{*}{Method} & \multirow{2}{*}{\begin{tabular}[c]{@{}c@{}}ACC\\ (Clean)\end{tabular}} & \multirow{2}{*}{\begin{tabular}[c]{@{}c@{}}ACC Drop\\ (Poisoning)\end{tabular}} \\
\multicolumn{2}{c|}{}                                                                                              &                         &                                                                        &                                                                         \\ \hline
\multicolumn{2}{c|}{\multirow{4}{*}{CIFAR10 / CIFAR100}}                                                           & Supervised              & 15.1\%                                                                 & 7.9\%                                                                   \\
\multicolumn{2}{c|}{}                                                                                              & SimCLR                  & 30.1\%                                                                 & 26.2\%                                                                   \\
\multicolumn{2}{c|}{}                                                                                              & BYOL                    & 32.1\%                                                                 & 23.4\%                                                                   \\
\multicolumn{2}{c|}{}                                                                                              & MoCo                    & 28.3\%                                                                 & 23.6\%                                                                   \\
\hline
\multicolumn{2}{c|}{\multirow{4}{*}{CIFAR100 / CIFAR10}}                                                           & Supervised              & 46.7\%                                                                 & 32.4\%                                                                  \\
\multicolumn{2}{c|}{}                                                                                              & SimCLR                  & 53.3\%                                                                 & 35.2\%                                                                  \\
\multicolumn{2}{c|}{}                                                                                              & BYOL                    & 60.7\%                                                                 &  44.3\%                                                                  \\
\multicolumn{2}{c|}{}                                                                                              & MoCo                    & 55.7\%                                                                 & 41.4\%                                                                 
\end{tabular}
\caption{Performance of fine-tuning on $\protect\sboth{\textsc{Trl}}{\mathrm{CL}}{\mathrm{fun}}$. \label{table:finetune}}}
\end{table}

\begin{mtbox}{Remark}
{\small The existing defenses are not easily retrofitted to defend against contrastive backdoor attacks in the setting of downstream tasks. }
\end{mtbox}

\section{Discussion}
\label{sec:discussion}

Thus far, we uncover the fundamental differences between supervised and contrastive backdoor attacks, propose possible explanations, and reveal the important implications entailed by such differences from a defense perspective. Next, we further explore a few key questions: (i) are our findings impacted by other factors (e.g., model architectures and  trigger definitions)? (ii) given the inadequacy of existing defenses, are there any promising alternatives? (iii) what are the limitations of this work?

\subsection{Other Factors}
\label{sec:factor}

We explore whether our findings are influenced by other factors, including model architectures and trigger definitions. 

\vspace{2pt}
{\bf Model Architectures --} To evaluate the influence of backbone model architectures, besides ResNet18, we also consider three other popular DNN architectures (inlcuding ResNet50\mcite{resnet}, ShuffleNet\mcite{shufflenet}, and MobileNet\mcite{mobilenetv2}) and train a SimCLR encoder on CIFAR10 as outlined in \msec{sec:comparison}.

\begin{figure}[t]
	\centering
	\includegraphics[width=1.0\columnwidth]{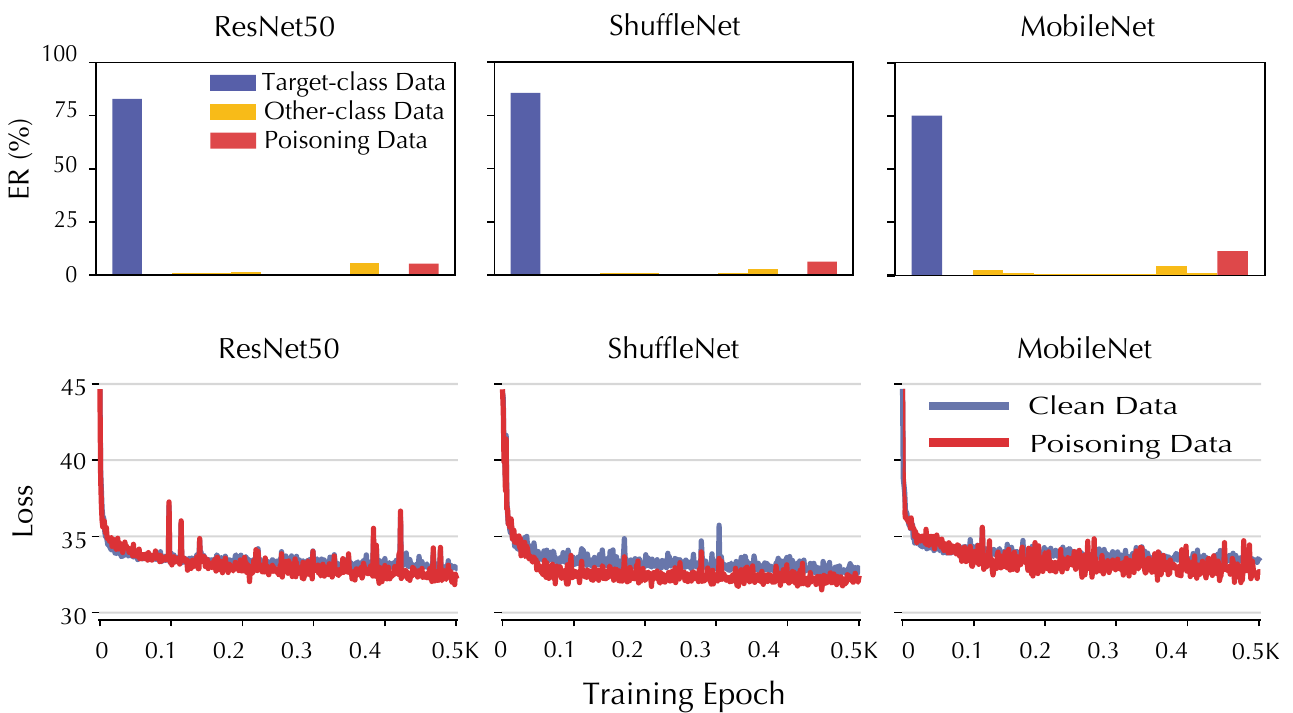}
	\caption{Learning dynamics and entanglement ratio under different backbone architectures. \label{fig:arch}}
\end{figure}

Figure\mref{fig:arch} shows the learning dynamics and entanglement ratio under varying backbone architectures. Our results indicate that: (i) the loss-decreasing pace is comparable for both poisoning and clean inputs across different architectures, and (ii) the poisoning inputs and the clean target-class inputs are entangled in the feature space under all the settings. The findings suggest that the backbone architectures have a limited impact on contrastive backdoor attacks.

\vspace{2pt}
{\bf Alternative triggers --}
In the previous evaluation, for contrastive attacks, we primarily focus on the functional trigger $\sboth{\textsc{Trl}}{\mathrm{CL}}{\mathrm{fun}}$. Here, we investigate whether the trigger definition impacts our findings. Specifically, we consider the universal trigger $\sboth{\textsc{Trl}}{\mathrm{CL}}{\mathrm{uni}}$ as an alternative trigger definition. Following\mcite{saha:2021:backdoor}, we specify a 5$\times$5 image patch as the trigger pattern, which is randomly placed at a random location of a given input.

\begin{figure}[!ht]
\centering
	\includegraphics[width=1.0\columnwidth]{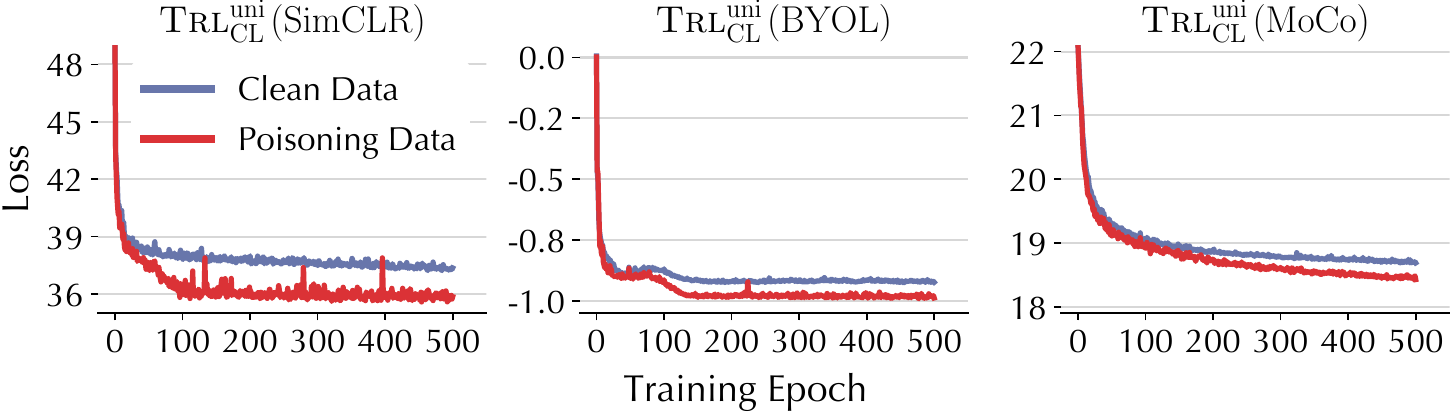}
    \caption{Learning dynamics of $\protect\sboth{\textsc{Trl}}{\mathrm{CL}}{\mathrm{uni}}$. \label{fig:bblbackdoor}}
\end{figure}

\begin{figure}[!ht]
	\centering
	\includegraphics[width=0.9\columnwidth]{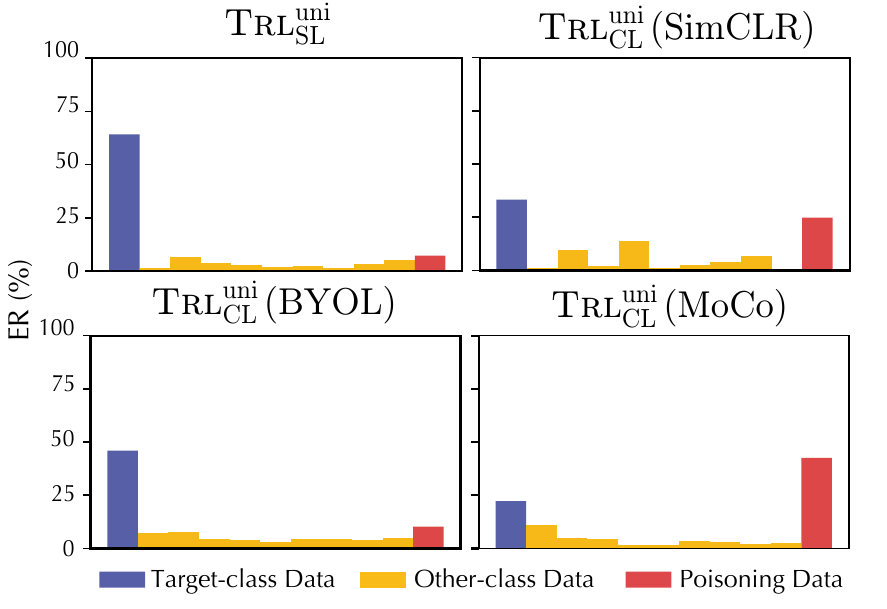}
	\caption{\rev{Entanglement ratios of $\protect\sboth{\textsc{Trl}}{\mathrm{SL/CL}}{\mathrm{uni}}$ on CIFAR10.}  \label{fig:er_uni}}
\end{figure}

Figure\mref{fig:bblbackdoor} shows the learning dyanmics of $\protect\sboth{\textsc{Trl}}{\mathrm{CL}}{\mathrm{uni}}$. Similar to $\sboth{\textsc{Trl}}{\mathrm{CL}}{\mathrm{fun}}$ (cf. Figure\mref{fig:leanring_dynamics}), $\sboth{\textsc{Trl}}{\mathrm{CL}}{\mathrm{uni}}$ also exhibits distinct learning dynamics compared with supervised attacks. Specifically, the contrastive loss of both poisoning and clean data decreases gradually at fairly similar paces. \rev{Further, Figure\mref{fig:er_uni} evaluates the entanglement ratios of $\sboth{\textsc{Trl}}{\mathrm{SL/CL}}{\mathrm{uni}}$. Observe that similar to Figure\mref{fig:er_distribution}, the entanglement effect is also more evident in the contrastive setting. Thus, the difference in entanglement effect is rather trigger-agnostic but stems from the underlying learning paradigms.}

\begin{figure}[t]
	\centering
	\includegraphics[width=0.9\columnwidth]{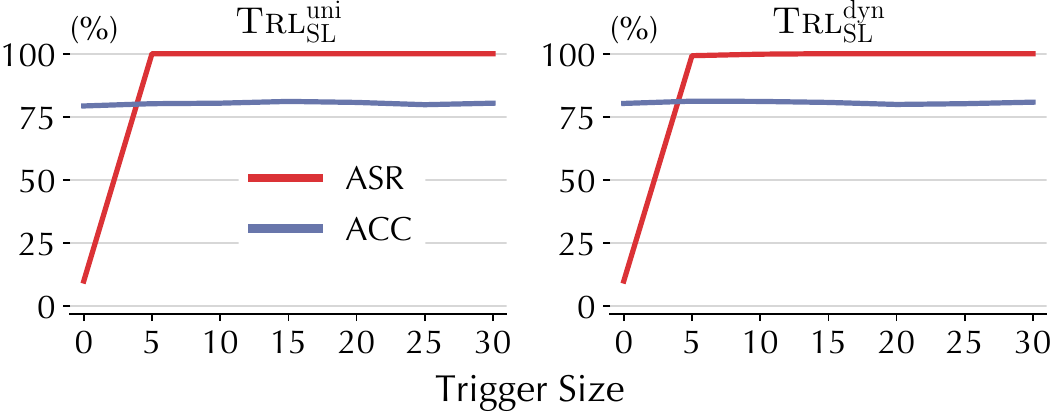}
	\caption{Attack performance of $\protect\sboth{\textsc{Trl}}{\mathrm{SL}}{\mathrm{uni}}$ and $\protect\sboth{\textsc{Trl}}{\mathrm{SL}}{\mathrm{dyn}}$ with respect to trigger strength (image patch size). \label{fig:magnitude_appendix}}
\end{figure}

Further, we conduct experiments to examine the influence of trigger strength (measured by the image patch size) on the effectiveness of supervised and contrastive attacks. Figure\mref{fig:magnitude_appendix} shows the ASR and ACC of $\sboth{\textsc{Trl}}{\mathrm{SL}}{\mathrm{uni}}$ and $\sboth{\textsc{Trl}}{\mathrm{SL}}{\mathrm{dyn}}$ as a function of trigger strength, which complements the results of Figure\mref{fig:overlap2}. Observe that in both cases of universal and dynamic triggers, as the trigger strength increases, the ASR of supervised attacks quickly peaks and remains around 100\%; meanwhile, the trigger strength has little impact on the ACC, indicating the independence of backdoor and benign tasks at the data level for supervised attacks.

\begin{table}[t]{\footnotesize 
\renewcommand{\arraystretch}{1.2}
\centering
\begin{tabular}{c|ccccc}
\multirow{2}{*}{CL Method} & \multicolumn{5}{c}{Trigger Size} \\ \cline{2-6} 
                        & 1    & 5    & 10   & 15   & 20   \\ \hline
SimCLR                  &  9.8\%    &  57.0\%    &  0.1\%    &  0.0    &  0.0    \\
BYOL                    &  10.7\%    &  34.6\%    &   65.6\%   & 2.01\%     & 0.3\%     \\
MoCo                    &  9.7\%    &  45.6\%    &   64.3\%   &  18.1\%    &   0.8\%  
\end{tabular}
\caption{Attack effectiveness of  $\protect\sboth{\textsc{Trl}}{\mathrm{CL}}{\mathrm{uni}}$ with respect to trigger size. \label{table:trigger_size}}}
\end{table}

In contrast,  as shown in Table\mref{table:trigger_size}, across different CL methods, as the trigger size increases from 1 to 20, the effectiveness of $\sboth{\textsc{Trl}}{\mathrm{CL}}{\mathrm{uni}}$ initially increases and then decreases abruptly, similar to the trend observed about  $\sboth{\textsc{Trl}}{\mathrm{CL}}{\mathrm{fun}}$ in Figure\mref{fig:overlap2}. 

\begin{mtbox}{Remark}
{\small The unique characteristics of contrastive backdoor attacks are agnostic to the concrete trigger definitions and backbone architectures.}
\end{mtbox}

\subsection{Potential Defenses}

We demonstrate that due to the specificities of contrastive backdoor attacks, defenses that attempt to segregate poisoning data based on learning dynamics and feature separability tend to fail. Given such limitations, we explore alternative defense strategies and discuss their potential challenges to defend against contrastive attacks.

\begin{figure}[t]
	\centering
	\includegraphics[width=0.75\columnwidth]{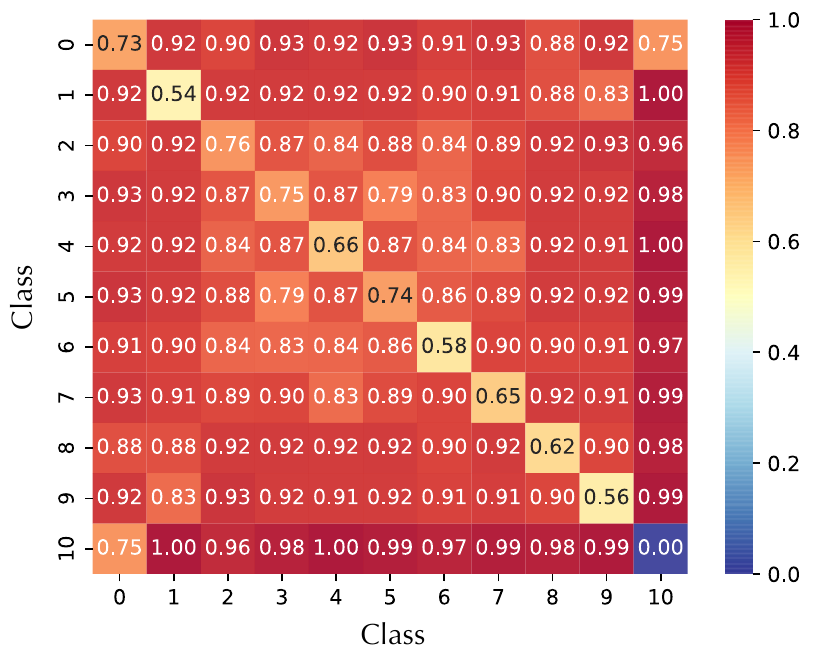}
	\caption{Average normalized pair-wise $L_2$ distance on CIFAR10 (trigger inputs: class 10; clean inputs: class 0 - 9). Diagonal cells represent intra-class distance while off-diagonal cells represent inter-class distance. \label{fig:dense_matrix}}
\end{figure}

\vspace{2pt}
{\bf Density-based Filtering --} We have an interesting observation in our evaluation:
the trigger inputs tend to form a cluster in the feature space, which is often much denser than the clusters formed by clean inputs. For example, we use a SimCLR model trained on CIFAR10 and measure the average pairwise $L_2$ distance between inputs in the feature space. Figure\mref{fig:dense_matrix} shows the normalized intra-class and inter-class distance. Observe that trigger inputs (class 10, with normalized distance 0.0) tend to cluster much more tightly compared with clean inputs (class 0 - 9, with normalized distance $\ge$ 0.56).

\rev{Motivated by this observation, we propose a density-based filtering defense. After training the encoder on potentially poisoning data, we use it to generate features for all inputs. Next, we apply a density-based clustering method (e.g., OPTICS\mcite{ankerst1999optics} or DBSCAN\mcite{ram2010density}) to cluster the training inputs based on their features. The densest clusters are identified as potentially poisoning data and subsequently removed from the training data. }

\begin{figure}[!ht]
	\centering
	\includegraphics[width=0.7\columnwidth]{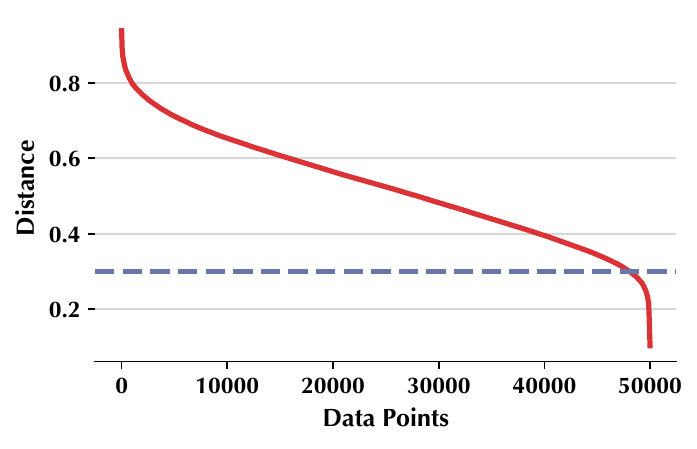}
	\caption{$K$-distance graph\label{fig:k_dis}}
\end{figure}

\rev{Below we evaluate this approach in defending against $\protect\sboth{\textsc{Trl}}{\mathrm{CL}}{\mathrm{fun}}$ (SimCLR) on CIFAR10 under two distinct poisoning rates: 1\% and 5\%. Specifically, we employ DBSCAN as the clustering algorithm to sift out poisoning inputs. DBSCAN hinges on two key parameters: (i) Minimum samples \(N_\mathrm{min}\), which is the minimum number of inputs required to form a cluster. Intuitively, a larger $N_\mathrm{min}$ results in denser clusters. Given that the number of poisoning inputs is typically small, we set $N_\mathrm{min}$ to either 30 or 100 in the evaluation. (ii) Maximum distance \(\epsilon\), which thresholds the distance between two inputs to determine whether they belong to the same cluster. To determine \(\epsilon\), we calculate the average distance between each input and its $K = N_\mathrm{min}$ nearest neighbors, plot the average $K$-distances in ascending order on a $K$-distance graph (as shown in Figure\mref{fig:k_dis}), and set $\epsilon$ as the point of maximum curvature, where the graph has the largest decreasing rate. We set $\epsilon = 0.3$ in the evaluation.}

\rev{Table\mref{table:density} summarizes the effectiveness of this density-based filtering measured by true positive rate (TPR) and false positive rate (FPR), as well as the attack success rate (ASR) and clean accuracy (ACC) of the model trained on the post-filtering training data. Notably, density-based filtering effectively sifts out a majority of poisoning inputs across various settings. For instance, under the setting of 5\% poisoning rate, $\epsilon = 0.3$, and $N_\mathrm{min} = 30$, it successfully filters 98.8\% of the poisoning inputs. Further, observe that after filtering potentially poisoning data and retraining the model, we may not only effectively mitigate the attack but also retain the model's utility. For instance, it achieves 13.3\% ASR and 80.1\% ACC under 1\% poisoning rate and $N_\mathrm{min} = 30$. However, the effectiveness of re-training seems sensitive to the parameter setting. For instance, under 1\% poisoning rate and $N_\mathrm{min} = 100$, as it fails to filter out approximately 100 poisoning inputs, the attack is not mitigated effectively.}

\rev{We further evaluate the effectiveness of density-based filtering with OPTICS as the underlying clustering algorithm, which is less sensitive to the parameter setting. It achieves an impressive 99.6\% TPR and 0\% FPR under 1\% poisoning rate. However, it has a much larger execution overhead. For example, on our platform, OPTICS requires 1.5 hours to complete the filtering, whereas DBSCAN finishes in approximately 10 seconds. Thus, we consider enhancing both the effectiveness and efficiency of density-based filtering defense as our ongoing work.}

\begin{table}[t]{\footnotesize 
\renewcommand{\arraystretch}{1.2}
\centering
\begin{tabular}{cc|c|cc|cc}
\multicolumn{2}{c|}{Parameter} & \multirow{2}{*}{Poisoning Rate} & \multicolumn{2}{c|}{Filtering} & \multicolumn{2}{c}{Re-training} \\
$\epsilon$                  & $N_\mathrm{min}$     &                               & TPR & FPR & ASR            & ACC            \\ \hline
\multirow{4}{*}{0.3}   & 30    & \multirow{2}{*}{1\%}          & 96.6\%    &   10.2\%                      &       13.3\%         &      80.1\%          \\
                       & 100   &                               & 80.1\%    &    0.4\%                      &   43.8\%             &    81.8\%            \\ \cline{3-7} 
                       & 30    & \multirow{2}{*}{5\%}          & 98.8\%     &  9.8\%                        &     19.9\%          &       77.8\%          \\
                       & 100   &                               & 97.8\%    &   0.4\%                       &    59.8\%            &  78.1\%            
\end{tabular}
\caption{Effectiveness of density-based filtering defense. \label{table:density}}}
\end{table}




\vspace{2pt}
{\bf Data-free pruning --} \rev{In addition to filtering potentially poisoning data, we also explore in-depth defenses for scenarios where filtering proves ineffective.}
We have shown that data-dependent pruning\mcite{liu2018fine} is insufficient to mitigate contrastive backdoor attacks, due to the indistinguishable feature patterns of clean and poisoning data (\msec{sec:feature-based-defense}). Surprisingly, data-free pruning, which is believed to be less effective than data-dependent pruning,  
shows promising performance against contrastive backdoor attacks.

\begin{figure}[!ht]
	\centering
	\includegraphics[width=0.9\columnwidth]{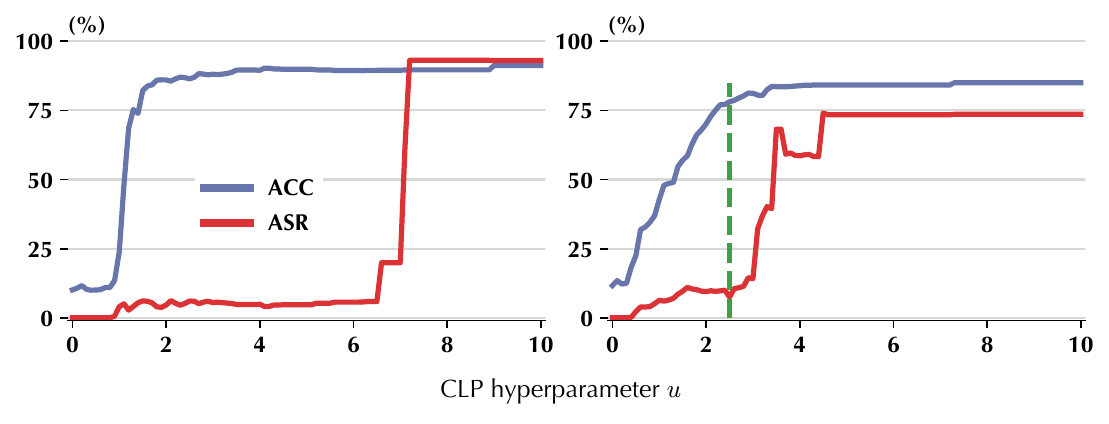}
	\caption{CLP against $\protect\sboth{\textsc{Trl}}{\mathrm{SL}}{\mathrm{fun}}$ and $\protect\sboth{\textsc{Trl}}{\mathrm{CL}}{\mathrm{fun}}$ on CIFAR10. \label{fig:clp_u}}
\end{figure}

Specifically, we consider channel Lipschitzness-based pruning (CLP)\mcite{datafree}, a data-free defense to remove backdoors. It exploits the observation that a subset of channels is more sensitive to trigger features compared to other channels, while the sensitivity of each channel can be estimated by the upper bound of its channel Lipschitz constant (UCLC), which can be computed in a data-free manner. Thus, given the mean ($\mu_k$) and variance ($\sigma_k$) of the UCLCs in the $k$-th convolutional layer, CLP prunes channels with UCLCs larger than $\mu_k + u\sigma_k$, where $u$ is a hyper-parameter. 

We apply CLP on models backdoored by $\sboth{\textsc{Trl}}{\mathrm{SL}}{\mathrm{fun}}$ and $\sboth{\textsc{Trl}}{\mathrm{CL}}{\mathrm{fun}}$ with varying $u$. As shown in Figure\mref{fig:clp_u}, by properly setting $u$, it is possible to effectively reduce the ASR of $\sboth{\textsc{Trl}}{\mathrm{CL}}{\mathrm{fun}}$ with limited ACC drop (about 5\%). However, compared with $\sboth{\textsc{Trl}}{\mathrm{SL}}{\mathrm{fun}}$, the proper range of $u$ for $\sboth{\textsc{Trl}}{\mathrm{CL}}{\mathrm{fun}}$ is much narrower ([2.2, 3.0] versus [1.8, 6.5]). 
\rev{To address this challenge, we propose to empirically set $u$ as the knee point of the ACC curve. Specifically, we measure the ACC under varying $u$, apply the Savitzky–Golay filter to smooth the curve\mcite{press1990savitzky}, and identify the knee point of the smoothed curve as the optimal $u$ during pruning. By applying this approach, we identify the optimal $u$ as 2.4 in the contrastive setting, as indicated by the green dashed line in Figure\mref{fig:clp_u}, which well balances utility and defensive efficacy (80.4\% ACC and 14.3\% ASR).}

\begin{figure}[!ht]
	\centering
	\includegraphics[width=0.7\columnwidth]{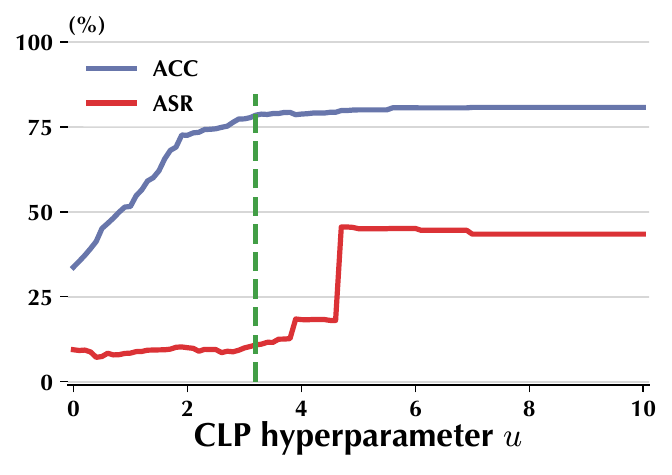}
	\caption{ CLP against the retrained model \label{fig:clp_density}}
\end{figure}

\vspace{2pt}
{\bf Ensemble defenses --} \rev{We further combine the previous two defenses to form a powerful ensemble defense against contrastive backdoor attacks. Specifically, we first apply density-based filtering to filter potentially poisoning data, retrain the model on the post-filtering training data, and then apply data-free pruning on the retrained model. We simulate the setting that the filtering hyperparameter is not properly set (e.g., $N_\mathrm{min} = 100$ under 1\% poisoning rate), which results in a retrained model with 43.8\% ASR and 81.8\% ACC. By applying data-free pruning on this retrained model, as illustrated in Figure\mref{fig:clp_density}, we further reduce ASR to 12.6\% while retaining ACC around 79.6\%. We believe this ensemble defense represents a promising direction worthy of further research.}


\subsection{Limitations and Future Work}

We further discuss the limitations of this work and point to several future directions. (i) {\bf Key properties:} our study primarily focuses on the differences between supervised and contrastive backdoor attacks, as reflected in learning dynamics and feature distributions. Other properties (e.g., neuron activation patterns) might also vary across these attack types. Future research could explore these underexplored properties to develop more effective defenses. (ii) {\bf Alternative defenses:}  our study highlights the defense implications of key differences between supervised and contrastive backdoor attacks. We illustrate the challenges faced by several representative defenses against contrastive backdoor attacks. Comprehensively investigating and benchmarking all the SOTA defenses (e.g.,\mcite{mna,rab}) against contrastive backdoor attacks is a valuable avenue for future research. (iii) {\bf Data modality:} \rev{while our study mainly focuses on the vision domain, contrastive learning has been extended to other domains such as natural language processing and multimodal learning\mcite{cl-text-generation,declutr}. It is worth investigating how domain constraints (e.g., discrete perturbation, semantic preservation, and interference between different modalities) affect contrastive backdoor attacks and their potential implications.} (iv) {\bf Generalization to self-supervised learning (SSL):} while focusing on contrastive learning methods, our findings might extend to the wider SSL paradigm (e.g., masked autoencoder\mcite{he2022masked}). It is essential, however, to validate these implications across broader contexts. Future work should explore the applicability and adaptability of our insights to alternative SSL frameworks and examine the implications for defenses in these new settings. 

In summary, while limited in several aspects, this work represents a solid step towards understanding and defending against the emerging threat of backdoor attacks in contrastive learning. Our findings highlight the need for defenses tailored to the specificities of contrastive backdoor attacks.

\section{Additional Related work}
\label{sec:literature}

In addition to the aforementioned related work, we further survey the literature most relevant to this work. 

\vspace{2pt}
{\bf Contrastive learning --} Recent years have witnessed the striding advances of contrastive learning (CL)\mcite{chen:2020:simple, grill:2020:bootstrap, chen:2021:exploring}. Compared with supervised learning (SL), CL obviates the reliance on data labeling and still learns high-quality representations from complex data, thereby facilitating various downstream tasks.  Meanware, the popularity of CL also spurs intensive research on its security properties.
Existing work explores the adversarial robustness of CL\mcite{jiang2020robust, fan2021does}. It is shown that, as a nice side effect, obviating the reliance on data labeling may benefit the model's robustness to adversarial examples, label corruption, and common data corruptions\mcite{ssl-robustness}. However, whether this robustness benefit also generalizes to other types of attacks remains an open question.

\rev{{\bf Supervised Contrastive learning --} Supervised contrastive learning (Sup-CL)\mcite{khosla2020supervised} introduces label information in the contrastive learning framework. By aligning latent representations with task-specific class semantics, Sup-CL enhances the quality of learned representations. 
Recent work\mcite{yang2021cade, chen2023continuous} applies Sup-CL in malware analysis and attains notable performance gains. Compared with Sup-CL, CL operates without explicit labels, relying on data augmentations to form and discern positive and negative pairs in an unsupervised manner.}\rev{Despite the differences, the attacks in this work can be easily adapted to binary classification (e.g., malware analysis) under both CL and Sup-CL settings. Specifically, by injecting the trigger pattern into a small number of goodware samples during training, both self-supervised and full-supervised variants of CL may inadvertently associate this pattern with goodware, as they seek to minimize the distance between similar entities in the latent space. At inference time, any malware with this pattern may be misclassified as goodware. We consider comparing the attacks under CL and Sup-CL settings in practical settings (e.g., malware analysis) as our ongoing work.}

\vspace{2pt}
{\bf Backdoor attacks --} As a major threat to machine learning security, backdoor attacks inject malicious backdoors into the victim's model during training and activate such backdoors at inference. Many backdoor attacks have been proposed for SL, which can be categorized along (i) attack targets -- input-specific\mcite{poisonfrog}, class-specific\mcite{tact} or any-input\mcite{badnet}, (ii) attack vectors -- polluting training data\mcite{trojannn} or releasing backdoored models\mcite{Ji:2018:ccsa}, and (iii) optimization metrics -- attack effectiveness\mcite{imc}, transferability\mcite{latent-backdoor}, or attack evasiveness\mcite{targeted-backdoor,poisonfrog,turner2019label,zhao2020clean}. 


Backdoor attacks are of particular interest for CL, potentially causing widespread damage due to their extensive application in downstream tasks. Supervised backdoor attacks, dependent on data labeling, are often inapplicable to CL, prompting innovative new approaches. For instance, BadEncoder\mcite{jia:2021:badencoder} injects backdoors into pre-trained encoders and releases backdoored models to victims, while SSLBackdoor\mcite{saha:2021:backdoor} uses image patch triggers to poison data; PoisonedEncoder\mcite{liu:2022:poisonedencoder} poisons the training data by randomly combining target inputs with reference inputs. Recently, CTRL\mcite{li2022demystifying} utilizes spectral triggers, achieving attack performance comparable to supervised attacks.

Yet, despite numerous studies on supervised and contrastive backdoor attacks, their fundamental differences remain unexplored. This work bridges this gap by revealing the distinctive mechanisms behind these attacks.


\vspace{2pt}
{\bf Backdoor defenses --} To mitigate the threats of backdoor attacks, many defenses have been proposed, which can be categorized according to their strategies\mcite{trojanzoo}: (i) input filtering, which purges poisoning examples from training data\mcite{tran:2018:nips,chen2018detecting}; (ii) model inspection, which determines whether a given model is backdoored and, if so, recovers the target class and the potential trigger\mcite{kolouri2020universal, huang2019neuroninspect, abs, Wang:2019:sp}; and (iii) input inspection, which detects trigger inputs at inference time\mcite{tact, gao2019strip, subedar2019deep}. 

However, mainly designed for supervised backdoor attacks, the effectiveness of these defenses in the CL setting remains unclear,  
raising several critical questions: can the existing defenses be retrofitted to CL attacks? If not, what new challenges do these attacks entail? How can we  address such challenges? This work systematically explores these key questions.

\section{Conclusion}

In this study, we examine the fundamental distinctions between supervised and contrastive backdoor attacks. Using a unified attack framework, we uncover that these attacks operate through different mechanisms, resulting in distinct learning dynamics and feature distributions. More importantly, we show that the unique characteristics of contrastive backdoor attacks entail important implications, requiring new and tailored defenses. Our findings shed new light on developing more robust contrastive learning techniques and point to several promising directions for further research.
\newpage





{\small \bibliographystyle{IEEEtran}
\bibliography{bibs/aml,bibs/optimization,bibs/general,bibs/ting,bibs/graph,bibs/ssl,bibs/interpretation}

\appendix

\section{\rev{Details of Contrastive Learning Methods}}
\label{sec:diff_cl}

\rev{Here, we compare the commonness and differences of representative contrastive learning methods.}

\rev{
SimCLR\mcite{chen:2020:simple} maximizes the similarity between two augmented versions of the same input ($x, x^+$) while minimizing the similarity amongst different inputs ($x, x^-$). The contrastive loss is defined by the InfoNCE loss:
\begin{equation}
-\log\frac{\exp \left(\frac{f_\theta(x)^\intercal  f_\theta(x^+)}{\tau}\right)}
      {
      \sum_{x^- \in \gN_{x}}  \exp\left(\frac{f_\theta(x)^\intercal  f_\theta(x^-)}{\tau}\right) + \exp\left(\frac{f_\theta(x)^\intercal  f_\theta(x^+)}{\tau}\right)
      },
\end{equation}
where \( f \) denotes the encoder function, \(\tau\) is a temperature parameter, the cosine similarity is used as the similarity metric, and \(\mathcal{N}_x\) represents $x$'s negative inputs.}



\rev{
BYOL\mcite{grill:2020:bootstrap} leverages two asymmetric networks: an online network, composed of an encoder \( f_\theta \) and a predictor \( q_\theta \), and a target network encompassing only the encoder \( f_\vartheta \) (parameterized by \( \vartheta \)). For each input, two augmented views \( (x, x^+) \) are generated and the contrastive loss is formulated as:
\begin{equation}
\label{eq:byol}
 \| q_\theta(f_\theta(x)) - f_\vartheta(x^+) \|_2^2
\end{equation}
where the representations are $L_2$ normalized. While \( \theta \) is optimized with respect to \meq{eq:byol}, \( \vartheta \) is updated as an exponential moving average of \( \theta \).}

\rev{
MoCo\mcite{he:2020:momentum} also employs two encoders, $f_{\theta_q}$ (query) and $f_{\theta_k}$ (key), to maximize the similarity between a query $x_q$ and its positive key $x_k^+$ while minimizing its similarity to negative keys. Given an input, its augmented pairs $(x_q, x^+_k)$ are considered as the query and its positive key, while other inputs are considered as its negative keys. The contrastive loss is defined as:
\begin{equation}
-\log\frac{\exp \left(\frac{f_{\theta_q}(x_q)^\intercal  f_{\theta_k}(x^+_k)}{\tau}\right)}
      {
      \exp \left(\frac{f_{\theta_q}(x_q)^\intercal  f_{\theta_k}(x^+_k)}{\tau}\right) + \sum_{x^-_k \in \mathcal{N}_{x_q}}  \exp\left(\frac{f_{\theta_q}(x_q)^\intercal  f_{\theta_k}(x^-_k)}{\tau}\right)
      },
\end{equation}
where $\tau$ is a temperature parameter and $\mathcal{N}_{x_q}$ is a set of negative keys. The key encoder $f_{\theta_k}$ is updated with a momentum-based approach.}

\rev{In conclusion, SimCLR and MoCo use both positive and negative pairs, albeit with different dictionary approaches in their contrastive loss formulations, while BYOL solely focuses on positive pairs. Meanwhile, MoCo and BYOL both leverage two encoders to stabilize the optimization, while SimCLR uses a single encoder.}

\section{Parameter Setting}
\label{sec:setting}

Table\mref{tab:hyper_training} lists the default parameter setting for training the encoder using different CL methods and SL models. 
Table\mref{tab:hyper_attack} lists the default parameter setting of different attacks in our evaluation.

\begin{table}[!ht]{\footnotesize
\centering
\renewcommand{\arraystretch}{1.1}
\begin{tabular}{c|cccc}
\multirow{2}{*}{Parameter} & \multicolumn{3}{c}{CL Method} & \multirow{2}{*}{SL}\\
\cline{2-4}
                                & SimCLR  & BYOL   & MoCo &\\ \hline
Optimizer                       & SGD     & SGD    & SGD   & SGD \\
Learning rate                    & 0.06     & 0.06    & 0.06 & 0.1\\
Optimizer momentum                    & 0.9     & 0.9    & 0.9 & 0.9\\
Momentum                       & -     & 0.996    & 0.999  &  - \\
Weight decay                    & 1e-4    & 1e-4   & 1e-4 &  1e-4 \\
Epochs                          & 500     & 500    & 500   &  20 \\
Batch size                      & 512     & 512    & 512  &   512\\
Temperature                     & 0.5     & -      & 0.5   &   - \\
Moving average                 & -       & 0.996  & -    & -  \\
Memory size                 & -       & -  & 65536    & -  \\
\end{tabular}
\caption{Parameter setting of encoder training. \label{tab:hyper_training} }}
\end{table}

\begin{table}[!ht]{\footnotesize
\centering
\setlength{\tabcolsep}{3pt}
\renewcommand{\arraystretch}{1.2}
\begin{tabular}{c|ccc}
\multirow{2}{*}{Parameter} & \multicolumn{3}{c}{Attack Type} \\
\cline{2-4}
                                & $\sboth{\textsc{Trl}}{\mathrm{CL}}{\mathrm{fun}}$ / $\sboth{\textsc{Trl}}{\mathrm{SL}}{\mathrm{fun}}$ & $\sboth{\textsc{Trl}}{\mathrm{CL}}{\mathrm{uni}}$ / $\sboth{\textsc{Trl}}{\mathrm{SL}}{\mathrm{uni}}$  &  $\sboth{\textsc{Trl}}{\mathrm{SL}}{\mathrm{dyn}}$\\ \hline
Poison ratio                       & 1\%    &  1\%    &  1\%     \\
Target class                    & 0   &  0    &  0    \\
Trigger size                   & -     & 5 $\times$ 5    & 32 $\times$ 32  \\
Magnitude                   & 100.0     & -   & -     \\
Block size                    & 32    & -   & -    \\
Frequency bands                         & 15, 31     & -    & -     \\
Trigger position              & -     & random     & 0.1     \\
Generator training epochs                     & -     & -      & 10      \\
Backdoor prob. $\rho_a$                  & -     & -      & 0.1     \\
Cross-trigger prob.  $\rho_b$                  & -     & -      & 0.1     \\

\end{tabular}
\caption{Default parameter setting of attacks. \label{tab:hyper_attack} }}
\end{table}

\section{Implementation Details}
\label{sec:impl}

\subsection{Different Variants of \attack}

We elaborate on the implementation of different variants of \attack.

$\sboth{\textsc{Trl}}{\mathrm{CL}}{\mathrm{fun}}$\mcite{li2022demystifying} --
$\mathsf{SelectCandidate}$ samples candidates from the target class.
$\mathsf{ApplyTrigger}$ applies a spectral trigger. It first converts the given input $x$ from the RGB space to the YC$_\text{b}$C$_\text{r}$ space. In the C$_b$ and C$_r$ channels, it divides $x$ into disjoint blocks (\meg, 32$\times$32). It applies discrete cosine transform (DCT)\mcite{rahman2013dwt} on each block to transform it from the spatial domain to the frequency domain. Then, it applies the pre-defined perturbation (increasing the magnitude by 50) on selected high-frequency bands. Then, it applies inverse DCT to transform the blocks back to the spatial domain and stitch them together. Finally, it converts the input from YC$_\text{b}$C$_\text{r}$ to RGB. 

$\sboth{\textsc{Trl}}{\mathrm{CL}}{\mathrm{uni}}$\mcite{saha:2021:backdoor} -- 
$\mathsf{SelectCandidate}$ samples candidates from the target class.
$\mathsf{ApplyTrigger}$ follows\mcite{saha:2021:backdoor}: it first generates a random 5$\times$5 image patch as the trigger and then applies it to a randomly selected position of the given candidate input.

$\sboth{\textsc{Trl}}{\mathrm{SL}}{\mathrm{fun}}$\mcite{li2022demystifying} -- $\mathsf{SelectCandidate}$ samples candidates from the target class. $\mathsf{ApplyTrigger}$ is the same as $\sboth{\textsc{Trl}}{\mathrm{CL}}{\mathrm{fun}}$. In addition, the labels of all the candidate inputs are set as the target class.

$\sboth{\textsc{Trl}}{\mathrm{SL}}{\mathrm{uni}}$\mcite{badnet,trojannn} -- $\mathsf{SelectCandidate}$ samples candidates from the target class. $\mathsf{ApplyTrigger}$ generates a 5$\times$5 image patch as the trigger and applies it to a random position of the given candidate. In addition, the labels of all the candidates are set as the target class. \rev{For a fair comparison, we let $\sboth{\textsc{Trl}}{\mathrm{SL}}{\mathrm{uni}}$ and $\sboth{\textsc{Trl}}{\mathrm{CL}}{\mathrm{uni}}$ share the same poisoning data.}

$\sboth{\textsc{Trl}}{\mathrm{SL}}{\mathrm{dyn}}$\mcite{nguyen2020input} -- $\mathsf{SelectCandidate}$ samples candidates across all the classes. $\mathsf{ApplyTrigger}$ generates input-specific triggers using a generative model $G$. Specifically, the training process runs in three modes: i) with probability $\rho_a \in (0, 1)$, it runs in the ``backdoor'' mode in which $G$ produces a trigger $G(x)$ and applied it to given input $x$, while its label is perturbed to the target class; ii) with probability $\rho_b \in (0, 1)$, it runs in the ``cross-trigger'' mode, in which $G$ produces the trigger $G(x)$ and applies it to a different input $x'$, while its original label is preserved; iii) with probability $1 - \rho_a - \rho_b$, it runs in the ``clean'' mode, in which $x$ is used as a clean input.



\subsection{Defense Details}

We detail the implementation of various defenses in our evaluation.

ABL\mcite{li2021anti} -- We follow the settings in their original paper. Specifically, we set the loss threshold $\gamma = 0.5$. If the loss of a training example goes below $\gamma$, gradient ascent will be activated to boost its loss to $\gamma$; otherwise, the loss stays the same. In the contrastive learning setting, we set the threshold $\gamma$ according to the same ratio compared to the largest loss.

AC\mcite{chen2018detecting} -- We set the hyper-parameters as the original paper.

STRIP\mcite{gao2019strip} -- In the original paper, it estimates the entropy distribution of clean samples on a validation set, selects an entropy threshold with a given FPR, and eventually removes all training samples with entropy below this threshold. In our work, we measure its effectiveness under three FPR settings: 0.5\%, 1.0\%, and 2.0\%.

SCAn\mcite{tact} -- Similar to STRIP\mcite{gao2019strip}, we vary the FPR under three settings: 0.5\%, 1.0\%, and 2.0\%,  and evaluate the TPR.

FP\mcite{liu2018fine} -- In the original paper, it keeps pruning the model until the tolerance of accuracy (5\%) reduction is reached. In our work, for a more detailed analysis, we show the ASR and ACC with respect to the variation of the pruned channels from 0 to 500.

\section{Additional Experiments}
\label{sec:addition}

\subsection{Attack effectiveness of $\protect\sboth{\textsc{Trl}}{\mathrm{CL}}{\mathrm{fun}}$}
In \msec{sec:factor}, we evaluate the universal trigger $\sboth{\textsc{Trl}}{\mathrm{CL}}{\mathrm{uni}}$ as an alternative trigger definition. Following\mcite{saha:2021:backdoor}, we specify a 5$\times$5 image patch as the trigger pattern, which is randomly placed at a random location of a given sample. Table\mref{table:asr_acc_addition} shows the clean accuracy and the corresponding attack success rate.

\begin{table}[!ht]{\footnotesize
\centering
\renewcommand{\arraystretch}{1.2}
 \setlength{\tabcolsep}{3pt}
 {
\begin{tabular}{cc|cccccc}
\multicolumn{2}{c|}{\multirow{3}{*}{Dataset}} & \multicolumn{3}{c}{Attack} \\ \cline{3-5} 
\multicolumn{2}{c|}{}                         &  \begin{tabular}[c]{@{}c@{}}\\ SimCLR\end{tabular} & \begin{tabular}[c]{@{}c@{}}$\sboth{\textsc{Trl}}{\mathrm{CL}}{\mathrm{fun}}$\\ BYOL\end{tabular} & \begin{tabular}[c]{@{}c@{}}\\ MoCo\end{tabular} \\ \hline
\multirow{2}{*}{CIFAR10}         & ASR (\%)        &   33.2 & 49.2 & 53.1  \\
                                 & ACC (\%)        &   79.4 & 84.3 & 80.6 \\
\end{tabular}}
\caption{Clean accuracy and attack success rate of $\protect\sboth{\textsc{Trl}}{\mathrm{CL}}{\mathrm{fun}}$. \label{table:asr_acc_addition}}}
\end{table}

\subsection{ABL against $\protect\sboth{\textsc{Trl}}{\mathrm{SL}}{\mathrm{uni}}$ and $\protect\sboth{\textsc{Trl}}{\mathrm{CL}}{\mathrm{uni}}$}
\label{sec:app_abl}

\rev{Table \ref{table:abl_uni} shows the effectiveness of ABL against $\protect\sboth{\textsc{Trl}}{\mathrm{SL}}{\mathrm{uni}}$ and $\protect\sboth{\textsc{Trl}}{\mathrm{CL}}{\mathrm{uni}}$. Observe that ABL is much less effective in the contrastive setting compared to the supervised setting.}

\begin{table}[!ht]\rev{{\footnotesize
\centering
\renewcommand{\arraystretch}{1.2}
\begin{tabular}{c|cc|cc}
\multirow{2}{*}{Isolation Rate} & \multicolumn{2}{c|}{$\sboth{\textsc{Trl}}{\mathrm{CL}}{\mathrm{uni}}$} & \multicolumn{2}{c}{$\sboth{\textsc{Trl}}{\mathrm{SL}}{\mathrm{uni}}$} \\ \cline{2-5} 
                                & TPR            & FPR            & TPR            & FPR           \\ \hline
1\%                             &    0.8\%       &  1.0\%     &    93.6\%          &      0.06\%        \\
5\%                             &     24.3\%      &    4.8\%     &    99.6\%          &     4.0\%          \\
10\%                            &   32.2\%        &   9.8\%     &    99.8\%            &     9.1\%        \\
20\%                            &    43.6\%      &    19.7\%     &    99.8\%         &    19.2\%         
\end{tabular}
\caption{\rev{ABL against $\protect\sboth{\textsc{Trl}}{\mathrm{SL}}{\mathrm{uni}}$ and $\protect\sboth{\textsc{Trl}}{\mathrm{CL}}{\mathrm{uni}}$ on CIFAR10.} \label{table:abl_uni}}}}
\end{table}

\subsection{STRIP against $\protect\sboth{\textsc{Trl}}{\mathrm{SL}}{\mathrm{uni}}$ and $\protect\sboth{\textsc{Trl}}{\mathrm{CL}}{\mathrm{uni}}$}
\label{sec:app_strip}

\rev{Table \ref{table:strip_uni} reveals a divergent defensive capacity of STRIP against distinct attacks: it encounters discernible difficulties when countering the contrastive backdoor attack, $\protect\sboth{\textsc{Trl}}{\mathrm{CL}}{\mathrm{uni}}$, as evidenced by a consistently low TPR at a designated FPR. In contrast, STRIP demonstrates a better defense against the supervised attack, $\protect\sboth{\textsc{Trl}}{\mathrm{SL}}{\mathrm{uni}}$, especially its dirty label variant, exhibiting a pronounced increase in TPR at a specified FPR. }

\begin{table}[!ht]\rev{{\footnotesize
\centering
\renewcommand{\arraystretch}{1.2}
\begin{tabular}{c|ccc}
Attack            & Decision  Threshold & FPR & TPR \\ \hline
\multirow{3}{*}{$\protect\sboth{\textsc{Trl}}{\mathrm{CL}}{\mathrm{uni}}$} &   5.11                  &   0.5\%  &   0.5\%  \\
                  &  5.4                   &   1.0\%  & 0.8\%    \\
                  &  5.5                  &  2.0\%   &   1.5\%  \\ \hline
\multirow{3}{*}{$\protect\sboth{\textsc{Trl}}{\mathrm{SL}}{\mathrm{uni}}$} &   4.17                  &   0.5\%  &   0.6\%  \\
                  &    5.3                &   1.0\%  &  2.4\%   \\
                  &    5.4                 &    2.0\% &   5.9\%  \\ \hline
\multirow{3}{*}{$\protect\sboth{\textsc{Trl}}{\mathrm{SL}}{\mathrm{uni}}$ (dirty)}  &  5.11                   &   0.5\%  &   0.6\%  \\
                  &   5.21                  &  1.0\%   & 30.6\%    \\
                  &   5.37                  &   2.0\%  &  59.8\%  
\end{tabular}
\caption{\rev{STRIP against $\protect\sboth{\textsc{Trl}}{\mathrm{SL}}{\mathrm{uni}}$ and $\protect\sboth{\textsc{Trl}}{\mathrm{CL}}{\mathrm{uni}}$ on CIFAR10.} \label{table:strip_uni}}}}
\end{table}

\subsection{Poisoning in Fine-tuning}

\rev{Figure\mref{fig:loss_fine} illustrates the learning dynamics of additional poisoning in the fine-tuning stage on CIFAR10. First, a backdoored encoder is obtained using \(\sboth{\textsc{Trl}}{\mathrm{SCL}}{\mathrm{fun}}\) (SimCLR). Subsequently, the encoder and the classifier head are fine-tuned on a small dataset, which includes 50 clean examples from each class, along with an additional 50 poisoning examples specifically from the target class. Throughout the fine-tuning stage, we continuously monitor the loss of both clean and poisoning data. In Figure\mref{fig:loss_fine}, we observe that the loss of poisoning data decreases much more rapidly than that of clean data, which is consistent with the supervised learning setting in \msec{sec:learn_dyn}. This is explained by that fine-tuning is inherently a form of supervised learning.}

\begin{figure}[!ht]
	\centering
	\includegraphics[width=0.7\columnwidth]{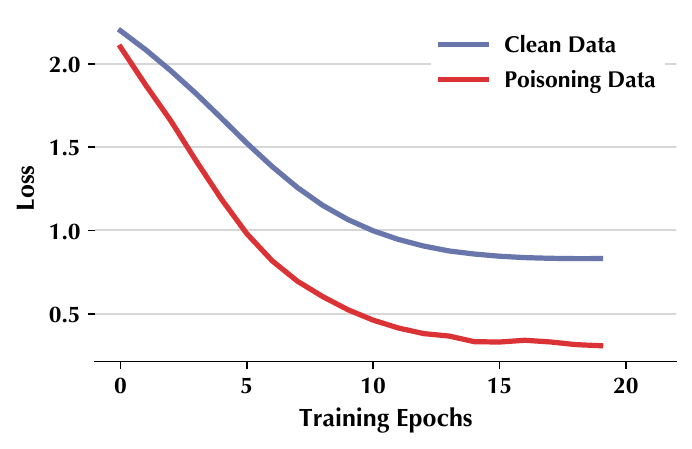}
	\caption{\rev{Learning dynamics of additional poisoning in fine-tuning.} \label{fig:loss_fine}}
\end{figure}




\end{document}